\UseRawInputEncoding
\documentclass[%
 reprint,
superscriptaddress,
 amsmath,amssymb,
 aps,
pra,
]{revtex4-2}

\usepackage{graphicx}
\usepackage{dcolumn}

\usepackage[english]{babel}		
\usepackage[utf8]{inputenc}
\usepackage[T1]{fontenc}
\usepackage{bbm}
\usepackage{amsthm}
\usepackage{amssymb}
\usepackage{amsmath}
\usepackage{graphicx}
\usepackage[nameinlink,capitalize]{cleveref}
\usepackage{booktabs}
\usepackage{multirow}

\crefname{fig}{Fig.}{Figs.}

\DeclareMathOperator{\trace}{Tr} 
\newcommand{\Tr}[1]{\trace\left[#1\right]}

\newcommand{\ket}[1]{|#1\rangle} 
\newcommand{\bra}[1]{\langle#1|} 
\newcommand{\braket}[2]{\langle #1|#2\rangle}
\newcommand{\ketbra}[2]{|#1\rangle\!\langle#2|}

\usepackage{soul}
\usepackage{xcolor}
\usepackage[normalem]{ulem}
\newcommand{\stkout}[1]{\ifmmode\text{\sout{\ensuremath{#1}}}\else\sout{#1}\fi}

\definecolor{textblue}{RGB}{20,120,165}
\definecolor{textterra}{RGB}{144,73,26}

\begin{document}

\title{Entangled Two-Photon Absorption in Cesium Atoms and the Limitations of the Far-Off-Resonance Approximation}

\author{Michael Caracas Núñez}
\affiliation{%
 Laboratorio de Óptica Cuántica, Universidad de los Andes, A.A. 4976, Bogotá, D.C., Colombia 
}
\author{Dario Egloff}
\affiliation{%
 Laboratorio de Óptica Cuántica, Universidad de los Andes, A.A. 4976, Bogotá, D.C., Colombia 
}
\affiliation{%
Escuela De Ciencias Exactas E Ingeniería,
Universidad Sergio Arboleda
Calle 74 No. 14-14 Bogotá
}%
 
\author{Mayerlin Nuñez Portela}%
 \email{m.nunez@uniandes.edu.co}
\affiliation{%
Laboratorio de Óptica Cuántica, Universidad de los Andes, A.A. 4976, Bogotá, D.C., Colombia 
}%

\date{\today}

\begin{abstract}
The discrepancies between theoretical and experimental results in the process of entangled two-photon absorption (ETPA) are not fully understood. Atomic systems are a promising alternative for investigating this process without the systematic effects present in molecules. We present a theoretical study of the ETPA process in cesium atoms, focusing on the $6^2S_{1/2} \rightarrow 8^2S_{1/2}$ transition. The ETPA cross section is evaluated both with and without the far-off-resonance (FOR) approximation, including the contributions from intermediate atomic states and decoherence effects. The quantum state of light considered is described by the joint spectral amplitude of photon-pairs produced by the spontaneous parametric down-conversion process. When the FOR approximation is applied, the enhancement factor is constant ($36\pi$). In contrast, without this approximation the enhancement factor oscillates with the entanglement time, and achieves a maximum value of $\sim 500$. These results show the limitations of approximations when calculating the ETPA cross section and contribute to the understanding of the discrepancies between theoretical and experimental values for the ETPA cross section in different samples. The numbers presented in this work can be a starting point for designing experiments aimed at measuring ETPA cross sections in alkali atoms.
\end{abstract}

\maketitle


\section{\label{sec:introduction}
Introduction
}

In the last decade, interest in understanding the interaction between quantum light and atoms or molecules has increased. Among the various phenomena in this domain, spectroscopic applications involving entangled photon-pairs have gained attention due to their unique quantum correlations and potential advantages over classical light sources. In particular, the two-photon absorption process induced by entangled photon pairs (ETPA) may be employed in microscopy, nonlinear spectroscopy, biological imaging, quantum information and sensing ~\cite{Goodson2024,dorfman2016nonlinear,schlawin2018entangled,gilaberte2019perspectives,szoke2020entangled,varnavski2020two,Bhai2025Entangled, schlawin2024two}. Most ETPA experiments have focused on measuring the ETPA cross section in molecular systems in solutions where the classical TPA process has been previously  studied~\cite{villabona2017entangled,Villabona-Monsalve2018,varnavski2020two,corona2022experimental,tabakaev2021energy,tabakaev2022spatial,he2024experimental}. The results of these experiments have been questioned when comparing the measured ETPA cross section, $\sigma_e$, with theoretical
values of the ETPA cross section calculated by means of a probabilistic approach~\cite{parzuchowski2021setting,hickam2022single,landes2024limitations}. In this model, the ETPA cross section is given by~\cite{Fei, landes2021quantifying}
\begin{equation}
	\sigma_e =F\frac{\sigma_c}{T_eA_0}. 
	\label{eq:Fei_model}
\end{equation}
Here, $\sigma_c$ is the classical TPA cross section, which can be obtained by driving the process with a CW laser. The characteristics of the entangled photons are quantified by the entanglement time, $T_e$, that corresponds to the correlation time of the photon pairs, and the entanglement area, $A_0$, that denotes the effective spatial overlap between the photon pairs and the atoms. The factor $F$ has been introduced as an enhancement factor for the ETPA process that depends on both the characteristics of the entangled photon-pair source and the parameters of the two-photon transition~\cite{landes2021quantifying}. In order to have an agreement between the experimental results on ETPA and those obtained from theory, it is necessary for $F$ to take values between $10^5$ and $10^{10}$ for molecular systems~\cite{hickam2022single}. Understanding the origin and magnitude of this factor is therefore crucial for resolving the discrepancies between the experimental measurements and calculations.

Theoretical models of entangled two-photon absorption contain approximations that can significantly impact the accuracy in the calculated ETPA cross section. Studies on molecules reveal that these approximations often overlook crucial physical mechanisms, such as vibrational and rotational modes~\cite{rodriguez2025perturbation} as well as permanent dipole moments~\cite{Burdick2018}, which play important roles in the dynamics of the ETPA process. When computing the ETPA cross section by means of the second-order perturbation theory, approximations on the spectrum of the photon-pairs and the so-called far-off-resonance (FOR) approximation are often applied, leading to an enhancement factor on the order of $F=1$~\cite{raymer_how_2021}. On the contrary, an enhancement factor higher than $1$ emerges when the spectrum of the quantum light is considered~\cite{drago2022aspects, rodriguez2025perturbation}.

In this work, we address the role of the FOR approximation in the enhancement factor for atomic systems when considering the spectrum of the quantum light. Atomic systems provide a promising alternative for testing theoretical models of ETPA due to their well-defined energy levels and the lack of interactions between atoms in vapor cells~\cite{dayan2004two, georgiades1995nonclassical}. These properties allow clearer observations of interference effects and entanglement-induced enhancements~\cite{dayan2007theory, landes2024limitations, Leon2013role, Du2011AtomicResonanceEnhanced}. We derive an analytical solution for the ETPA cross section considering the explicit form of the frequency correlations of the entangled photon-pairs produced in a spontaneous parametric down-conversion process (SPDC). In particular, we explore the feasibility of measuring ETPA in cesium by theoretically determining the ETPA cross section for the $6^2S_{1/2}\to 8^2S_{1/2}$ transition. The energy of this transition can be achieved by means of photon-pairs produced by commercial non-linear crystals pumped at $411$~nm. Experimental values for the classical TPA cross section in this transition have been previously reported~\cite{CaracasNunez:23}. 

Section~\ref{sec:basics} of this article presents the basic theory to describe a two-photon absorption process considering the interaction of atoms with a quantized electric field in the density operator formalism. The description of the quantum state of entangled photon pairs is presented in section~\ref{sec:state_of_biphoton}. With this information, the cross section of the ETPA process for the $6^2S_{1/2}\to 8^2S_{1/2}$ transition in cesium atoms is calculated by employing the FOR approximation and without it in section IV. The effects of the FOR approximation in the enhancement factor are explicitly addressed in section~\ref{sec:discussion}. These results serve as a foundation for the design and optimization of future experiments aiming to observe ETPA in Cesium or similar alkali atoms.

\section{\label{sec:basics}
Theory of the TPA Cross Section}

To compute the TPA cross section, we consider a hydrogen-like atomic system with a ground state $\ket{g}$, a final excited state $\ket{f}$ and intermediate states collectively denoted by $\ket{e}, \ket{e'}$. Each state has a corresponding energy eigenvalue $E_l = \hslash\omega_l$ ($l=g,e,e',f$). The system interacts with photon-pairs from a radiation field. The cross section of the TPA process can be obtained from the probability to excite a transition from $\ket{g}$ to $\ket{f}$ using these two photons.

The time evolution of the complete system, atom plus photons, can be studied by means of the density operator $\hat{\rho}(t)$. The individual systems are independent before the interaction, therefore, the initial state can be expressed as the tensor product $\hat{\rho}_0=\hat{\rho}_A(0)\otimes\hat{\rho}_F(0)$, where $\hat{\rho}_A(0)$ corresponds to the atomic initial state, and $\hat{\rho}_F(0)$ represents the initial state of the photons. In the interaction picture ($I$), $\hat{\rho}(t)$ satisfies the Von Neumann equation:
\begin{equation}
    \frac{\partial \hat{\rho}(t)}{\partial t} = \frac{1}{i\hslash}\left[ \hat{H}_{I}(t), \hat{\rho}(t) \right].
    \label{eq:tempevolution}
\end{equation}

The Hamiltonian that describes this interaction, $\hat{H}_{I}(t)$, is
\begin{equation}
    \hat{H}_{I}(t) = \hat{\mathbf{d}}_I(t)\cdot\hat{\mathbf{E}}_I(t),
    \label{eq:H_interaction}
\end{equation}
where the electric field, $\hat{\mathbf{E}}_I(t)=\hat{\mathbf{E}}^{(+)}(t) + \hat{\mathbf{E}}^{(-)}(t)$, is described by photons propagating along the $z$ direction with a given spectral distribution. As calculated explicitly in Appendix~\ref{sec:AppendA},
\begin{equation}
    \hat{\mathbf{E}}^{(+)}(z,t) = i\int_{-\infty}^{\infty}\frac{\mathrm{d}\omega}{2\pi}\left( \frac{\hslash\omega}{2\epsilon_0cA_0} \right)^{1/2}\hat{a}(\omega)e^{i\left[k(\omega)z-\omega t\right]}\mathbf{e}.
    \label{eq:quantized_EM4}
\end{equation}

Here, $\mathbf{e}$ is the polarization of the photons, the operators $\hat{a}(\omega)$ are responsible for the annihilation of a photon with a given frequency $\omega$, $c$ is the speed of light, $\epsilon_0$ is the vacuum permittivity, $A_0$ is the transverse area of the photons, and $k(\omega)$ is the corresponding wave number. From Eq.~\eqref{eq:quantized_EM4}, it is clear that $ \hat{\mathbf{E}}^{(-)}(z,t) = \left[ \hat{\mathbf{E}}^{(+)}(z,t) \right]^{\dagger}$.

In order to describe the atom-photon interaction, it is necessary to introduce the real matrix elements of the electric dipole operator, defined as $d_{jk}=d_{kj}^*=d_{kj}=\bra{j}\mathbf{d}\cdot\mathbf{e}\ket{k}$, with $\mathbf{d}=e\mathbf{r}$ as the electric dipole moment. Therefore, the electric dipole operator in the interaction picture is given by $\hat{d}_I(t)=\hat{d}_I^{(+)}(t)+\hat{d}_I^{(-)}(t)$, where, 
\begin{align}\label{eq:dipole_operator}
    \hat{d}_I^{(-)}(t) &=\sum_{k,j|j>k}d_{jk}\ket{j}\bra{k}e^{i(\omega_j-\omega_k)t}, \\  
    \hat{d}_I^{(+)}(t) &= \left[\hat{d}_I^{(-)}(t) \right]^\dagger=\sum_{k,j|j<k}d_{jk}\ket{j}\bra{k}e^{-i(\omega_j-\omega_k)t}.\notag
\end{align}
        
The indexes of the sum of Eq.~\eqref{eq:dipole_operator} run over all possible eigenstates of the atomic system, $\ket{j}$, $\ket{k}$, with frequencies $\omega_j$ and $\omega_k$ respectively. 

For the TPA process, the time evolution of the state, $\hat{\rho}(t)$, is given by a fourth-order expansion~\cite{Raymer2021}:
\begin{equation}
    \begin{matrix}
        \hat{\rho}(t) &=\left( \frac{1}{i\hslash} \right)^{4}\sum_{p,q,r,s}\int_{-\infty}^{t}dt_4\int_{-\infty}^{t_4}dt_3\int_{-\infty}^{t_3}dt_2\int_{-\infty}^{t_2}dt_1 \\ 
	&\times \left[ \hat{H}_{I}^{(s)} 
        (t_4),\left[\hat{H}_{I}^{(r)}(t_3),\left[ \hat{H}_{I}^{(q)}(t_2),\left[ \hat{H}_I^{(p)}(t_1),\hat{\rho}_0
	\right] \right] \right] \right].
    \end{matrix}
    \label{eq:rhosolution}
\end{equation}

The sum is performed over $p,q,r,s$, corresponding to all the possible combinations of $(\hat{d}^{(+)}+\hat{d}^{(-)})(\hat{E}^{(+)}+\hat{E}^{(-)})$. Using the rotating-wave approximation, only terms of the form $\hat{H}_{I}^{(p)}=\hat{d}^{(-p)}\hat{E}^{(p)}$, with $p=\pm$ will survive. Starting from $\hat{\rho}(t)$, the TPA probability, $p_{g\rightarrow f}$, for the atom to perform the $\ket{g}\to \ket{f}$ transition is given by
\begin{equation}
    p_{g\rightarrow f}=\Tr{\hat{\rho}(t)\ket{f}\bra{f}}.
    \label{eq:TPAprobabilidad}
\end{equation}

The trace is calculated over the complete system. Note that the only terms that contribute to the TPA probability are those of the form $\hat{H}_{I}^{(+)}(t_a)\hat{H}_{I}^{(+)}(t_b)\hat{\rho}_0\hat{H}_{I}^{(-)}(t_c)\hat{H}_{I}^{(-)}(t_d)$. Different time orderings of these operators lead to three distinct quantum pathways for the $\ket{g}\to \ket{f}$ transition: the double quantum coherence (DQC) path where $t_a=t_4,~t_b=t_3,~t_c=t_1$ and $t_d=t_2$, the non-rephasing path (NRP) where $t_a=t_4,~t_b=t_2,~t_c=t_1$ and $t_d=t_3$, and the rephasing path (RP, $t_a=t_3,~t_b=t_2,~t_c=t_1$ and $t_d=t_4$)~\cite{Raymer2021}. The contribution of the DQC path is dominant when the sum of the frequencies of the photon-pairs is near-resonant to the atomic two-photon transition and neither of the individual photons is in resonance with intermediate transitions~\cite{Raymer2021}. The NRP and RP paths contribute strongly when the difference between the frequencies of the photon-pairs is comparable to the transition frequencies of the intermediate states~\cite{drago2022aspects}. For the problem discussed in this manuscript, we consider the two photons to be in resonance with the two-photon transition, and each individual photon to be off-resonance with the intermediate states. Therefore, the DQC term dominates in the transition probability, and Eq.~\eqref{eq:TPAprobabilidad} becomes
\begin{widetext}
    \begin{equation}
        p_{g\rightarrow f}=\frac{1}{\hslash^{4}}\int_{-\infty}^{t}dt_4\int_{-\infty}^{t_4}dt_3\int_{-\infty}^{t_3}dt_2\int_{-\infty}^{t_2}dt_1 \trace\left[ \hat{H}_{I}^{(+)}(t_4)\hat{H}_{I}^{(+)}(t_3)\hat{\rho}_0\hat{H}_{I}^{(-)}(t_1)\hat{H}_{I}^{(-)}(t_2) \right] + c.c.
        \label{eq:TPAprobability1}
    \end{equation}
\end{widetext}

In order to compute this expression, we consider that at $t=0$, the initial state, $\hat{\rho}_0$, is separable in the atomic and field parts, and the fact that the field and the dipole operators commute, $[\hat{d}_I^{(\pm)}(t),\hat{E}^{(\pm)}(t)]=0$. Therefore, the trace in Eq.~\eqref{eq:TPAprobability1} can be written as the product 
\begin{equation}
    \small
    \trace\left[ \hat{H}_{I}^{(+)}(t_4)\hat{H}_{I}^{(+)}(t_3)\hat{\rho}_0\hat{H}_{I}^{(-)}(t_1)\hat{H}_{I}^{(-)}(t_2) \right] = C_AC_F.
    \label{eq:correlation_function_2}
\end{equation}
where
\begin{align}
     \small
    C_A &=\trace\left[ \hat{d}^{(-)}(t_4)\hat{d}^{(-)}(t_3)\hat{\rho}_A(0)\hat{d}^{(+)}(t_1)\hat{d}^{(+)}(t_2)\ket{f}\bra{f} \right ].
    \label{eq:CM_1} \\
      \small
    C_F &=\trace\left[ \hat{E}^{(+)}(t_4)\hat{E}^{(+)}(t_3)\hat{\rho}_F(0)\hat{E}^{(-)}(t_1)\hat{E}^{(-)}(t_2) \right ].
    \label{eq:correlation_function}
\end{align}

Here $C_A$ is the trace over the atomic subsystem representing an atomic correlation function and the trace of the radiation field, $C_F$, is the second-order field correlation function~\cite{loudon2000quantum}.

In order to calculate $C_A$ it is necessary to define the initial state $\hat{\rho}_A(0)=\ket{g}\bra{g}$, and to describe different decoherence mechanisms acting on the atoms. In particular, the line shape of an atomic transition is a random process that describes the interaction between the atoms and the environment. As the environment is not included in this theoretical description, this interaction leads to a decoherence effect. The Kubo theory allows to include such decoherence mechanism by adding an imaginary term, $i\gamma_{jk}$, to the transition frequencies, such that $\omega_{jk}=\omega_j-\omega_k$ becomes $\omega_{jk}+i\gamma_{jk}$~\cite{Raymer2021, kubo1969stochastic}. Here, $\gamma_{jk}$ is approximated by the linewidth of the transition $\ket{k}\to\ket{j}$~\cite{Raymer2021}. Appendix~\ref{append:C} shows the expression of $C_A$ in terms of the transition frequencies, given the following change of variables $r=t_4-t_3$, $s=t_3-t_2$ and $\tau=t_2-t_1$. As a result, the atomic correlation function can be written as  
\begin{equation}
    \small
    C_A =\sum_{e,e'}D_{fe'eg}e^{-(\gamma_{fe'}-i\omega_{fe'})r}e^{-(\gamma_{fg}-i\omega_{fg})s}e^{-(\gamma_{eg}-i\omega_{eg})\tau},
    \label{eq:atomcorrelation}
\end{equation}
where $D_{fe'eg} = d_{fe'}d_{e'g}d_{ge}d_{ef}$ is the product of the different dipole moments.   

The second-order field correlation function, $C_F$, can be calculated considering an initial state $\hat{\rho}_F(0)=\ketbra{\Psi}{\Psi}$, with the condition that the final state of the atomic system is $\ket{f}\bra{f}$. Therefore, $C_F$ can be written as  
\begin{equation}
    \small
    C_F = \bra{\Psi}\hat{E}^{(-)}(t_1)\hat{E}^{(-)}(t_2)\ket{0}\bra{0}\hat{E}^{(+)}(t_3)\hat{E}^{(+)}(t_4)\ket{\Psi},
    \label{eq:CF_1}
\end{equation}
where $\ket{0}$ is the vacuum state and $\ket{\Psi}$ describes a two-photon state driving the TPA process. The probability of the TPA process can be calculated from Eq.~\eqref{eq:TPAprobability1} with the correlation functions in Eqs.~\eqref{eq:atomcorrelation} and~\eqref{eq:CF_1}, such that
\begin{equation}
    p_{g\rightarrow f}=\frac{1}{\hslash^{4}}\int_{-\infty}^{t}dt_4\int_{-\infty}^{t_4}dt_3\int_{-\infty}^{t_3}dt_2\int_{-\infty}^{t_2}dt_1 C_F C_A + c.c.
    \label{eq:TPAprobability1.2}
\end{equation}

For a TPA process driven by entangled photon-pairs, the rate (probability per unit of time) is linear with the photon flux~\cite{javanainen1990linear}. This dependence will be addressed explicitly in the following sections. Therefore, the ETPA cross section can be obtained from the probability in Eq.~\eqref{eq:TPAprobability1.2} as
\begin{equation}
   \sigma_e=\frac{p_{g\to f}}{T\phi},
   \label{eq:sigmae_prime1.1}
\end{equation}
where $\phi$ is the flux of photon pairs and $T$ is the time of interaction. In order to calculate this cross section explicitly, first we must describe the quantum state of the entangled photon-pairs involved in the second-order field correlation function in Eq.~\eqref{eq:CF_1}.

\section{State of the Biphoton}
\label{sec:state_of_biphoton}

The state of the entangled photon-pairs is described by the state of a biphoton produced via the SPDC process. This process is a nonlinear interaction where a pump beam with frequency $\omega_p$ impinges on a nonlinear crystal with second-order electric susceptibility represented by the tensor $\chi^{(2)}$, producing two down-converted photons. These photons are historically known as \textit{signal} and \textit{idler}, with frequencies $\omega_s$ and $\omega_i$, respectively \cite{boyd}. The process satisfies the so-called phase-matching conditions $\omega_p=\omega_s+\omega_i$ and $\vec{k_p}=\vec{k_s}+\vec{k_i}$~\cite{YanhuaShih_2003, couteau2018spontaneous}. The Hamiltonian of this process can be written as~\cite{YanhuaShih_2003}
\begin{equation}
    \small
    \hat{H}_{SPDC}(t)=\epsilon_0\int_{\mathcal{V}}d^{3}\mathbf{r} \chi^{(2)}\hat{E}_p^{(+)}(\mathbf{r},t)\hat{E}_s^{(-)}(\mathbf{r},t)\hat{E}_i^{(-)}(\mathbf{r},t) + h.c.
    \label{eq:Hamil_SPDC}
\end{equation}

Here, $\mathcal{V}$ is the interaction volume and $\hat{E}_p^{(+)}$ is considered as a classical electromagnetic monochromatic plane wave propagating in the $z$ direction. The fields for the signal and idler photons are described according to Eq.~\eqref{eq:quantized_EM4}. The contribution of the transverse spatial degrees of freedom are not taken into account since we consider that the \textit{signal} and \textit{idler} photons propagate in collinear configuration with the pump beam. Therefore, the resulting two-photon state is a superposition of the frequency modes of the signal and idler photons, as explained in Appendix~\ref{append:B}1. With this, it is possible to write the biphoton state as~\cite{couteau2018spontaneous,YanhuaShih_2003}
\begin{equation}   
    \small
    \ket{\Psi} = \varepsilon\int_{-\infty}^{\infty}\mathrm{d}\omega_s\int_{-\infty}^{\infty}\mathrm{d}\omega_i\hat{a}^{\dagger}(\omega_s)\hat{a}^{\dagger}(\omega_i)\Phi(\omega_s,\omega_i)\ket{0},
    \label{eq:biphoton_state3}
\end{equation}
where $\Phi(\omega_s,\omega_i)$ is the Joint-Spectral Amplitude (JSA) of the photon pairs and $\varepsilon$ is the probability amplitude of producing down-converted photons.  

Here, we consider a type-II SPDC process driven with a monochromatic pump laser. This process generates two photons. The central frequency of each photon, $\omega_0$, is related to the frequency of the pump laser light by $\omega_p=2\omega_{0}$. We can define the detunings from $\omega_0$ for the signal and idler photons as $\omega_{s,i}=\omega_0 \pm \Omega_{s,i}$. Considering a small detuning, $\Omega_{s,i}\ll\omega_0$, the wave numbers $k(\omega)$ in the nonlinear crystal can be approximated by a Taylor series around $\omega_0$. Truncating the series at first order, we get $k_{s,i}(\omega_{s,i}) \approx k_{s,i}(\omega_0) \pm k'_{s,i}(\omega_0)\Omega_{s,i}$~\cite{YanhuaShih_2003}. With this expression and the phase-matching conditions it follows that $\Omega_s=\Omega_i$ for a monochromatic pump beam. As detailed in Appendix~\ref{append:B}2, the explicit expression for the JSA for a type-II SPDC process is
\begin{align}
    \Phi(\omega_{s},\omega_i)&=\Phi(\Omega_s+\omega_0,\omega_0-\Omega_i)\\
  &= \sqrt{\frac{T_e}{2\pi}}f(\Omega_s,\Omega_i)\left(  \frac{\varphi(\Omega_s,\Omega_i)+\varphi(\Omega_i,\Omega_s)}{2} \right),
	\label{eq:type-II-JSA2}  
\end{align}
where $f(\Omega_s,\Omega_i)$ is the spectral distribution of the pump light, and
\begin{equation}
    \varphi(\Omega_s,\Omega_i) = \text{sinc}\left[ \left( (k_s'-k_p')\Omega_s+(k_p'-k_i')\Omega_i\right)l/2 \right].
\end{equation}

In Eq.~\eqref{eq:type-II-JSA2}, $T_e=(k_s'-k_i')l$ is the so-called entanglement time of the down-converted photons, which depends on the length, $l$, of the nonlinear crystal and the group velocities of the photons, $k_s'$ and $k_i'$~\cite{YanhuaShih_2003}. In Appendix~\ref{append:B}3, it is shown that the JSA function must be symmetric under exchange of variables.

Once the state of the field is defined, it is possible to calculate the second-order field correlation function in Eq.~\eqref{eq:CF_1}. To do so, we replace the initial state of the field by the state of the biphoton in Eq.~\eqref{eq:biphoton_state3} and the fields of the signal and idler photons (Eq.~\eqref{eq:quantized_EM4}). With these considerations, $C_F$ takes the form (see Appendix~\ref{append:B}3)
\begin{align}
	C_F &= \kappa\int_{-\infty}^{\infty}\mathrm{d}\omega_s'\int_{-\infty}^{\infty}\mathrm{d}\omega_i'\int_{-\infty}^{\infty}\mathrm{d}\omega_s\int_{-\infty}^{\infty}\mathrm{d}\omega_i|\varepsilon|^2 \notag \\
    &\times\Phi^{*}(\omega_s',\omega_i')\Phi(\omega_s,\omega_i)e^{-i\omega_s'\tau}e^{-i(\omega_s + \omega_i)s}e^{-i\omega_sr},
	\label{eq:fieldcorrelation2}
\end{align}
where $\kappa=\left(\frac{\hslash\omega_0}{\epsilon_0cA_0}\right)^2$ and $|\varepsilon|^2$ is related to the flux of entangled-photon pairs. In the following section, we employ this result to calculate the ETPA cross section from the probability to drive a two-photon transition with these entangled photon-pairs.

\section{ETPA cross section in Cesium atoms}

The probability of the ETPA process is calculated by replacing the atomic and field correlation functions (Eqs.~\eqref{eq:atomcorrelation} and~\eqref{eq:fieldcorrelation2}) in Eq~\eqref{eq:TPAprobability1.2}. We performed the time integrals following the procedure in Appendix~\ref{append:E}1. The result can be summarized as
\begin{equation}
    p_{g\to f} = \frac{|\varepsilon|^2 \kappa}{\hslash^4}\sum_{e,e'}D_{fe'eg} I_{e',e},
    \label{eq:TPAprobability2}
\end{equation}
where $I_{e',e}$ is
\begin{widetext}
\begin{equation}
I_{e,e'}=\int_{-\infty}^{\infty}\mathrm{d}\omega_s'\int_{-\infty}^{\infty}\mathrm{d}\omega_s\int_{-\infty}^{\infty}\mathrm{d}\omega_i\frac{\Phi^{*}(\omega_s',\omega_i+\omega_s-\omega_s')\Phi(\omega_s,\omega_i)}{[\gamma_{fe'}+i(\omega_i - \omega_{fe'})]\left[\gamma_{fg}+i(\omega_s+\omega_i-\omega_{fg})\right][\gamma_{eg} +i(\omega_s' - \omega_{eg})]} + c.c.
\label{eq:brrr2}
\end{equation}
\end{widetext}

Here, we used the phase-matching condition $\omega_p=\omega_s+\omega_i=\omega_s'+\omega_i'$, to reduce the number of integrals in frequency. The ETPA cross section can therefore be expressed from Eq.~\eqref{eq:sigmae_prime1.1}, taking into account that the flux of photon pairs produced by the SPDC process is given by $\phi=2|\varepsilon|^2/A_0T$. Therefore,
\begin{equation}
   \sigma_e = \frac{\kappa A_0}{2\hslash^4}\sum_{e,e'}D_{fe'eg} I_{e',e}.
   \label{eq:sigmae_prime}
\end{equation}

The value of the ETPA cross section for the $6^2S_{1/2} \rightarrow 8^2S_{1/2}$ transition in cesium atoms is obtained by evaluating the integrals in Eq.~\eqref{eq:brrr2}. For this we employ two methods: First, using the usual far-off-resonance approximation \cite{Raymer2021, landes2021quantifying, drago2022aspects}, second, calculating analytically the values of the cross section without the FOR approximation. With these two methods we aim to understand the role of this approximation in the enhancement factor $F$ in Eq.~\eqref{eq:Fei_model}. For both cases, we considered the JSA as given by Eq.~\eqref{eq:type-II-JSA2}.

\subsection{\label{subsec:FOR} ETPA cross section with the far-off-resonance approximation}

In order to understand the FOR approximation, we define  $\Omega_{fe'} = \omega_0 - \omega_{fe'}$ and $\Omega_{eg} = \omega_0 - \omega_{eg}$ as the detunings of the frequency of the down-converted photons with respect to the frequencies of the intermediate transitions. When these detunings are much higher than $\Omega_i$ and $\Omega_s'$, it is possible to approximate $\omega_i - \omega_{fe'}\approx \omega_0 - \omega_{fe'} = \Omega_{fe'}$ and $\omega_s' - \omega_{eg}\approx \omega_0 - \omega_{eg} = \Omega_{eg}$. The FOR approximation also includes the fact that, for atomic systems, $\gamma_{fe'}\ll\Omega_{fe'}$ and $\gamma_{eg}\ll\Omega_{eg}$, such that $\gamma_{fe'}+i(\omega_i - \omega_{fe'})\approx i\Omega_{fe'}$ and $\gamma_{eg}+i(\omega_{s'} - \omega_{eg})\approx i\Omega_{eg}$. With these considerations, it is possible to simplify Eq.~\eqref{eq:brrr2}, as
\begin{align}
    I_{e,e'} &= \frac{1}{(-\omega_{fe'}+\omega_0)(\omega_{eg}-\omega_0)}\int_{-\infty}^{\infty}d\omega_s'\int_{-\infty}^{\infty}d\omega_s\int_{-\infty}^{\infty}d\omega_i \notag \\
    &\times \frac{\Phi^{*}(\omega_s',\omega_i+\omega_s-\omega_s')\Phi(\omega_s,\omega_i)}{\gamma_{fg}+i(\omega_s+\omega_i-\omega_{fg})} + c.c.
	\label{eq:integrales}
\end{align}

For the joint spectral amplitude in a type-II SPDC process  (Eq.~\eqref{eq:type-II-JSA2}), it is possible to obtain an analytical solution for this integral (see Appendix~\ref{append:H}). The result is
\begin{equation}
    I_{e,e'} = \frac{4\pi}{T_e}\frac{1}{(-\omega_{fe'}+\omega_0)(\omega_{eg}-\omega_0)}\frac{\gamma_{fg}}{\gamma_{fg}^2 + \Delta^2},
\end{equation}
with $\Delta=2\omega_0-\omega_{fg}$ as the detuning of the pump frequency ($2\omega_0$) of the SPDC process with respect to the frequency of the two-photon transition. Therefore, the ETPA cross section, according to Eq.~\eqref{eq:sigmae_prime}, is 
\begin{equation}
    \small
    \sigma_e^{(2)} = \frac{2\pi A_0\kappa}{T_e\hslash^4} \sum_{e,e'}\frac{D_{fe'eg}}{(-\omega_{fe'}+\omega_0)(\omega_{eg}-\omega_0)}\frac{\gamma_{fg}}{\gamma_{fg}^2 + \Delta^2}.
    \label{eq:sigmae_FOR1}
\end{equation}

This ETPA cross section is related to the classical TPA cross section, $\sigma_c(\omega_0)$, calculated using the second-order perturbation theory, by~\cite{CaracasNunez:23}
\begin{equation}
    \sigma_e (\omega_0)=\frac{36\pi}{T_eA_0}\sigma_c(\omega_0). 
    \label{eq:sigmae/sigma}
\end{equation}

The dependence of the ETPA cross section with $T_e$ and $A_0$ is expected from the probabilistic model in Eq.~\eqref{eq:Fei_model}, obtaining an enhancement factor of $F=36\pi$.

\begin{figure}[ht]
    \centering \includegraphics[width=1.0\linewidth]{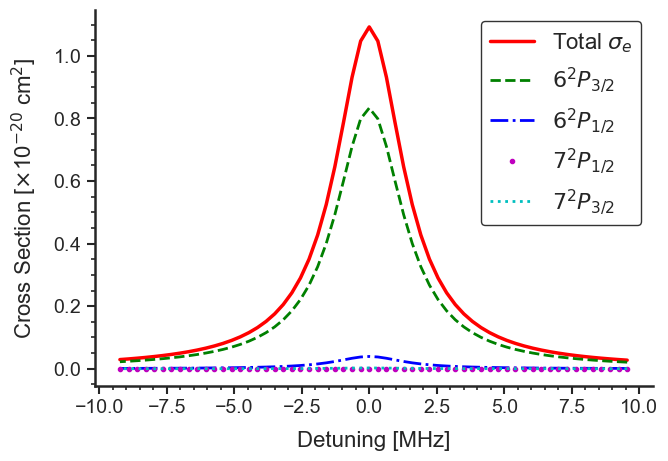}
    \caption{Far-off-resonance ETPA cross section as a function of the detuning, $\Delta/2\pi$, of the pump beam with respect to the two-photon transition $6^2S_{1/2}\to 8^2S_{1/2}$ in cesium atoms. The solid line corresponds to the total ETPA cross section, i.e, considering all the spectroscopic information in Appendix E. Dashed lines are the individual cross sections for each of the intermediate states. This graphic is obtained considering an entanglement time of $T_e = 100$ fs and an entanglement area of $A_0=1.8\times 10^{-8}$~m$^2$.}
    \label{fig:FOR_CS_Type_II}
\end{figure}

The cross section for the $6^2S_{1/2} \to 8^2S_{1/2}$ transition is evaluated simulating the JSA of photon-pairs generated by a $l=0.5$~mm BBO crystal corresponding to an entanglement time of $T_e=100$~fs. The non-linear crystal interacts with a monochromatic CW pump laser light at $411$~nm. The entanglement area considered for the photon-pairs in this calculation is $A_0=1.8\times 10^{-8}$~m$^2$, corresponding to a waist of $75$~$\mu$m. Figure~\ref{fig:FOR_CS_Type_II} shows the behavior of $\sigma_e$ as a function of the detuning $\Delta$. All the relevant spectroscopic information of cesium atoms required to obtain a value of $\sigma_e$ is presented on Appendix~\ref{sec:cs_values}. The contributions of each individual path $6^2S_{1/2} \to \ket{e} \to 8^2S_{1/2}$ are illustrated as dashed lines. It is observed in each of the curves the Lorentzian profile characteristic of the natural line-shape of the atomic transition. The most important contribution to the total ETPA cross section comes from the transition path through the $6^2P_{3/2}$ state. This is expected since the contribution of the intermediate state $\ket{e}$ to the cross section becomes smaller when the detunings $\Omega_{fe'}$ and $\Omega_{eg}$ are bigger. The $7P$ energy levels have higher detunings from the central frequency of the entangled photons, so their contributions are much smaller than those of the $6P$ states. Other dipole-allowed transitions will have higher detunings, therefore, their contributions can be neglected~\cite{CaracasNunez:23}. The main result of this calculation is that, at resonance, i.e. $\Delta=0$, the maximum value for the ETPA cross section is $\sigma_e = 1.1\times 10^{-20}$~cm$^2$.

\subsection{\label{subsec:without_FOR} ETPA cross section without the far-off-resonance approximation}

In this section, the analytical values of $\sigma_e$ without performing the FOR approximation in Eq.~\eqref{eq:brrr2} are presented. Here, we still considered the photon-pairs to be off-resonant with the intermediate states of the atom. Following the Appendix~\ref{append:G}, it is possible to write $I_{e,e'}$ for a type-II SPDC process as the sum of four terms such that
\begin{equation}
    I_{e,e'} = \frac{T_e}{\pi}(I_{e,e'}^{(1)}+I_{e,e'}^{(2)}+I_{e,e'}^{(3)}+I_{e,e'}^{(4)}),
\end{equation}
and the ETPA cross section becomes 
\begin{equation}
    \small
    \sigma_e (\omega_0,T_e) = \frac{\kappa A_0}{2\hslash^4}\frac{T_e}{\pi}\sum_{e,e'}D_{fe'eg}(I_{e,e'}^{(1)}+I_{e,e'}^{(2)}+I_{e,e'}^{(3)}+I_{e,e'}^{(4)}),
    \label{eq:sigmae_full}
\end{equation}
where
\begin{equation}
    I_{e,e'}^{(1)} = \frac{\gamma_{fe'}\gamma_{fg}\gamma_{eg}}{\gamma_{fg}^2 + \Delta^2}S_1(\Omega_{fe'}, \gamma_{fe'}, T_e) S_1(-\Omega_{eg}, \gamma_{eg}, T_e),
    \label{eq:R1_}
\end{equation}
\begin{equation}
     I_{e,e'}^{(2)} = \frac{\gamma_{eg}\Delta}{\gamma_{fg}^2 + \Delta^2} S_2(\Omega_{fe'}, \gamma_{fe'}, T_e) S_1(-\Omega_{eg}, \gamma_{eg}, T_e),
     \label{eq:R2_}
\end{equation}
\begin{equation}
    I_{e,e'}^{(3)} = -\frac{\gamma_{fe'}\Delta}{\gamma_{fg}^2 + \Delta^2}S_1(\Omega_{fe'}, \gamma_{fe'}, T_e)S_2(-\Omega_{eg}, \gamma_{eg}, T_e),
    \label{eq:R3_}
\end{equation}
\begin{equation}
    I_{e,e'}^{(4)} = \frac{\gamma_{fg}}{\gamma_{fg}^2 + \Delta^2} S_2(\Omega_{fe'}, \gamma_{fe'}, T_e)S_2(-\Omega_{eg}, \gamma_{eg}, T_e),
    \label{eq:R4_}
\end{equation}
and the auxiliary functions, $ S_1(\Omega_{jk}, \gamma_{jk}, T_e)$ and $S_2(\Omega_{jk}, \gamma_{jk}, T_e)$, are defined by
\begin{align}
    S_1(\Omega_{jk}, \gamma_{jk}, T_e) &= \int_{-\infty}^{\infty}\frac{\text{sinc}[(\Omega + \Omega_{jk})T_e/2]}{(\gamma_{jk}^2 + \Omega^2)}\mathrm{d}\Omega, \\
    S_2(\Omega_{jk}, \gamma_{jk}, T_e) &= \int_{-\infty}^{\infty}\frac{\Omega\text{sinc}[(\Omega + \Omega_{jk})T_e/2]}{(\gamma_{jk}^2 + \Omega^2)}\mathrm{d}\Omega,
\end{align}
with the new variables $\Omega= \Omega_s - \Omega_{fe'}$ and $\Omega' = \Omega'_s + \Omega_{eg}$. The explicit form of the integrals $S_1$ and $S_2$ are (see Appendix~\ref{append:G})
\begin{widetext}
    \begin{equation}
        S_1(\Omega_{fe'}, \gamma_{fe'}, T_e) = \left(\frac{2\pi}{T_e}\right)\frac{\gamma_{fe'} - e^{-\gamma_{fe'}T_e/2} \left[ \gamma_{fe'} \cos \left(\frac{\Omega_{fe'}T_e}{2}\right)-\Omega_{fe'} \sin \left(\frac{ \Omega_{fe'}T_e}{2}\right)\right]}{\gamma_{fe'} \left(\gamma_{fe'}^2+\Omega_{fe'}^2\right)},
        \label{eq:oscillations_1}
    \end{equation}
    \begin{equation}
        S_2(\Omega_{fe'}, \gamma_{fe'}, T_e) = \left(\frac{2\pi}{T_e}\right)\frac{-\Omega_{fe'} +  e^{-\gamma_{fe'}T_e/2} \left[ \gamma_{fe'} \sin \left(\frac{\Omega_{fe'}T_e}{2}\right)+\Omega_{fe'} \cos \left(\frac{ \Omega_{fe'}T_e}{2}\right)\right]}{ \left(\gamma_{fe'}^2+\Omega_{fe'}^2\right)}.
       \label{eq:oscillations_2}
    \end{equation}
\end{widetext}

Replacing these expressions in Eqs.~(31)-(34), the ETPA cross section is obtained as a function of the entanglement time $T_e$. The result exhibits damped oscillations as a function of $T_e$ and the detunings $\Omega_{jk}$. Summing over the contributions of the intermediate states in Eq.~\eqref{eq:sigmae_full}, it is possible to observe that the cross section becomes a sum of a product of two damped oscillations. This generates an interference pattern with a rich structure when the cross section is calculated for different experimental parameters, e.g., $T_e$ and $\Delta$.  

\begin{figure}[ht]
    \centering
    \includegraphics[width=1.0\linewidth]{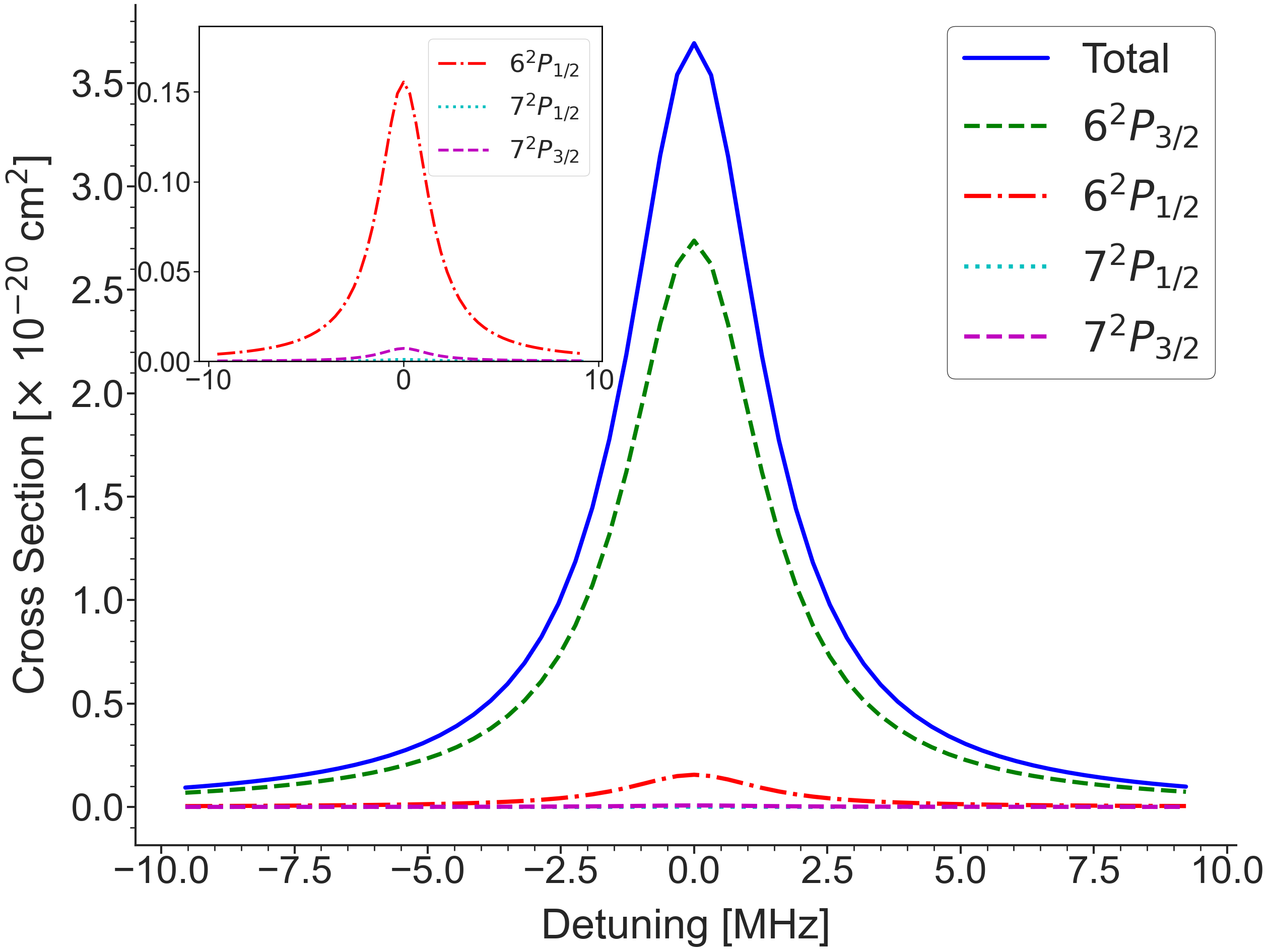}
    \caption{ETPA cross section without the far-off-resonance approximation as a  function of the detuning $\Delta/2\pi$ for the TPA transition $6^2S_{1/2}\to 8^2S_{1/2}$ in cesium atoms. The inset shows the cross section for the intermediate transition paths with the lowest contributions to the total ETPA cross section.}
    \label{fig:CS_complete_Te_Fixed}
\end{figure}

The behavior of the ETPA cross section as a function of the detuning $\Delta$ is presented in Fig.~\ref{fig:CS_complete_Te_Fixed} for a fixed value of $T_e=100$~fs. The values are obtained for the same  type-II SPDC source considered in the FOR approximation.  The contributions of each of the four paths considered in the FOR approximation are presented as dashed lines with their corresponding Lorentzian profiles.  The solid curve is associated to the total ETPA cross section in Eq.~\eqref{eq:sigmae_full}. Similarly to the FOR approximation, the most important contribution to the total ETPA cross section comes from the transition path through the $6^2P_{3/2}$ state. Nevertheless, the maximum value for the $\sigma_e$ in this case is $\approx 3.4$ times higher than the one obtained in the FOR approximation for the same $T_e$.  

\begin{figure}[h!]
    \centering
    \includegraphics[width=1.0\linewidth]{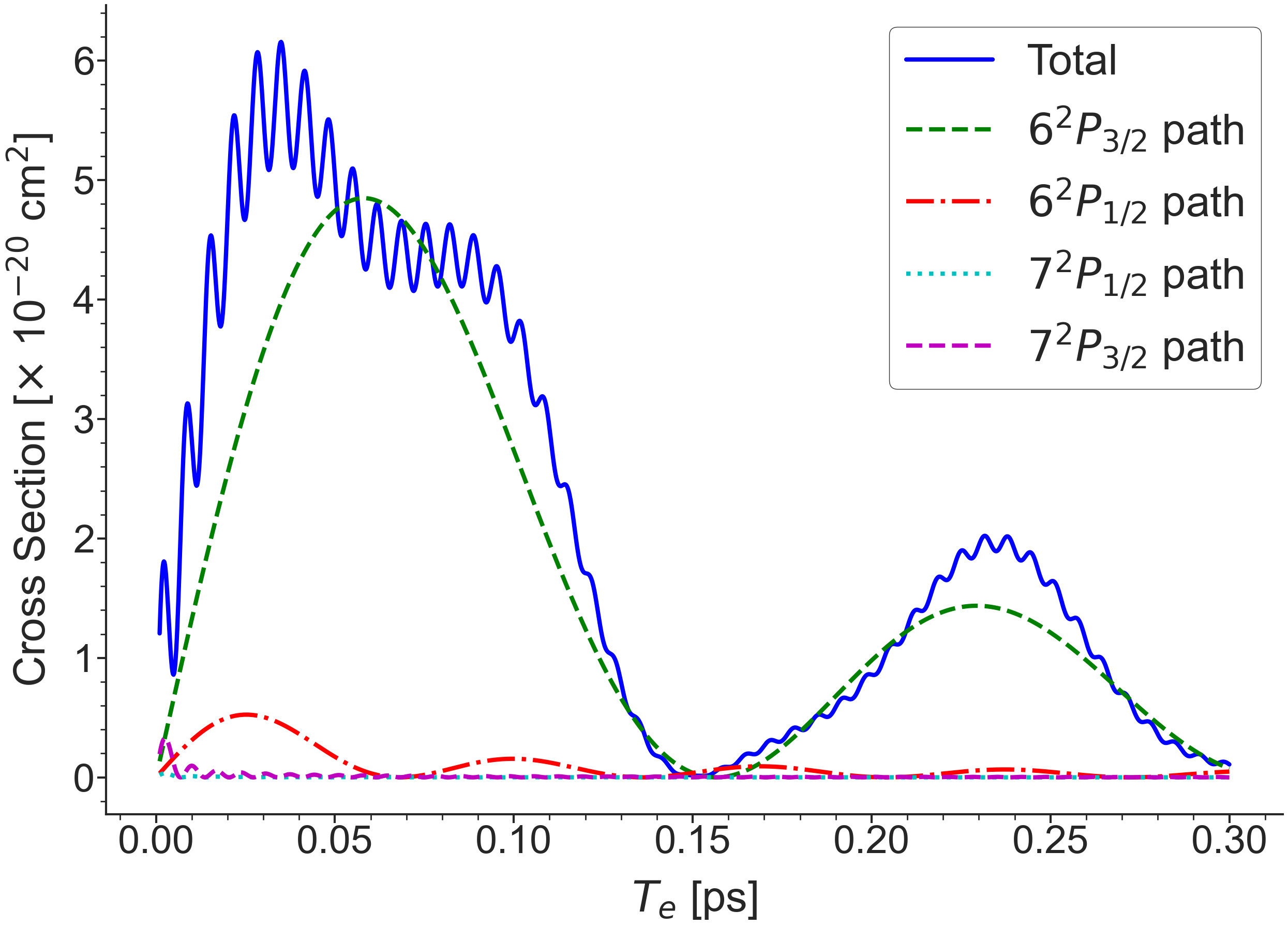}
    \caption{ETPA cross section without the far-off-resonance approximation as a function of the entanglement time $T_e$ for $\Delta=0$ (resonance) for the TPA transition $6^2S_{1/2}\to 8^2S_{1/2}$ in cesium atoms. The inset shows the cross section for the intermediate transition paths with the lowest contributions to the total ETPA cross section.}
    \label{fig:CS_complete_D0_Fixed}
\end{figure}

The behavior of the ETPA cross section as a function of the entanglement time $T_e$ is presented in Fig.~\ref{fig:CS_complete_D0_Fixed} for the resonance condition, $\Delta=0$. Five curves are displayed. The dashed lines correspond to the ETPA cross section computed for each individual transition path. The solid line shows the ETPA cross section computed with the 4 intermediate states considered. Damped oscillations are observed in the curves, as expected from Eqs. (37) and (38). The total ETPA cross section can be interpreted  as an interference pattern that depends on the relative phases ($\Omega_{jk}$) accumulated by photons traversing different intermediate states. As expected, the main contribution comes from the path through the $6^2P_{3/2}$ state. However, there are values of $T_e$ where the contributions of this path are higher than the total cross section, indicating destructive interference between the different pathways at these specific times.  Constructive interference can also be observed for values of $T_e$ where the solid line is much higher that the dashed line for the $6^2P_{3/2}$ state.

Table~\ref{tab:results2} illustrates the effects of constructive interference for $T_e=100$~fs. In particular, it shows the individual contributions of each intermediate transition path to the cross section and the total $\sigma_e$ calculated according to Eq.~(30). The detunings between $\omega_0$ and the intermediate transition frequencies $\omega_{eg}$ are presented on resonance with the two-photon transition ($\Delta=0$). The value of the last cross section is higher than the algebraic sum of the individual contributions, showing the importance of the interference effects to determine the enhancement factor. These observations can provide a mechanism for controlling the ETPA efficiency through temporal engineering of the entangled photon-pairs source~\cite{Burdick2018, Leon2013role}.

\begin{table}[ht]
    \centering
    \begin{tabular}{ccc}
        \toprule
        \textbf{Transition Path} & \textbf{$\sigma_e$ (cm$^2$)} & $|\Omega_{eg}|$\textbf{(THz)} \\
        \midrule
        $6^2P_{1/2}$ & $1.6 \times 10^{-21}$ & 29.39 \\
        $6^2P_{3/2}$ & $2.7 \times 10^{-20}$ & 12.78 \\
        $7^2P_{1/2}$ & $9.6 \times 10^{-24}$ & 288.52 \\
        $7^2P_{3/2}$ & $7.1 \times 10^{-23}$ & 293.61 \\\hline
        Total algebraic sum & $2.9 \times 10^{-20}$ \\\hline
        Total (Eq. \eqref{eq:sigmae_full}) & $3.7 \times 10^{-20}$ \\
        \bottomrule
    \end{tabular}
    \caption[Theoretical results obtained for the ETPA cross-section considering the contribution of the intermediate states]{Theoretical results obtained for the ETPA cross section in resonance computed for each individual transition path without the FOR approximation. The algebraic sum of these values is compared with the total ETPA cross section according to Eq. \eqref{eq:sigmae_full}. The detunings, $\Omega_{eg}$, of the central frequency of the entangled photon-pairs and the intermediate atomic transition are also presented.}
    \label{tab:results2}
\end{table}

\section{\label{sec:discussion} Discussion}

\begin{figure}[h!]
    \centering
    \includegraphics[width=1.0\linewidth]{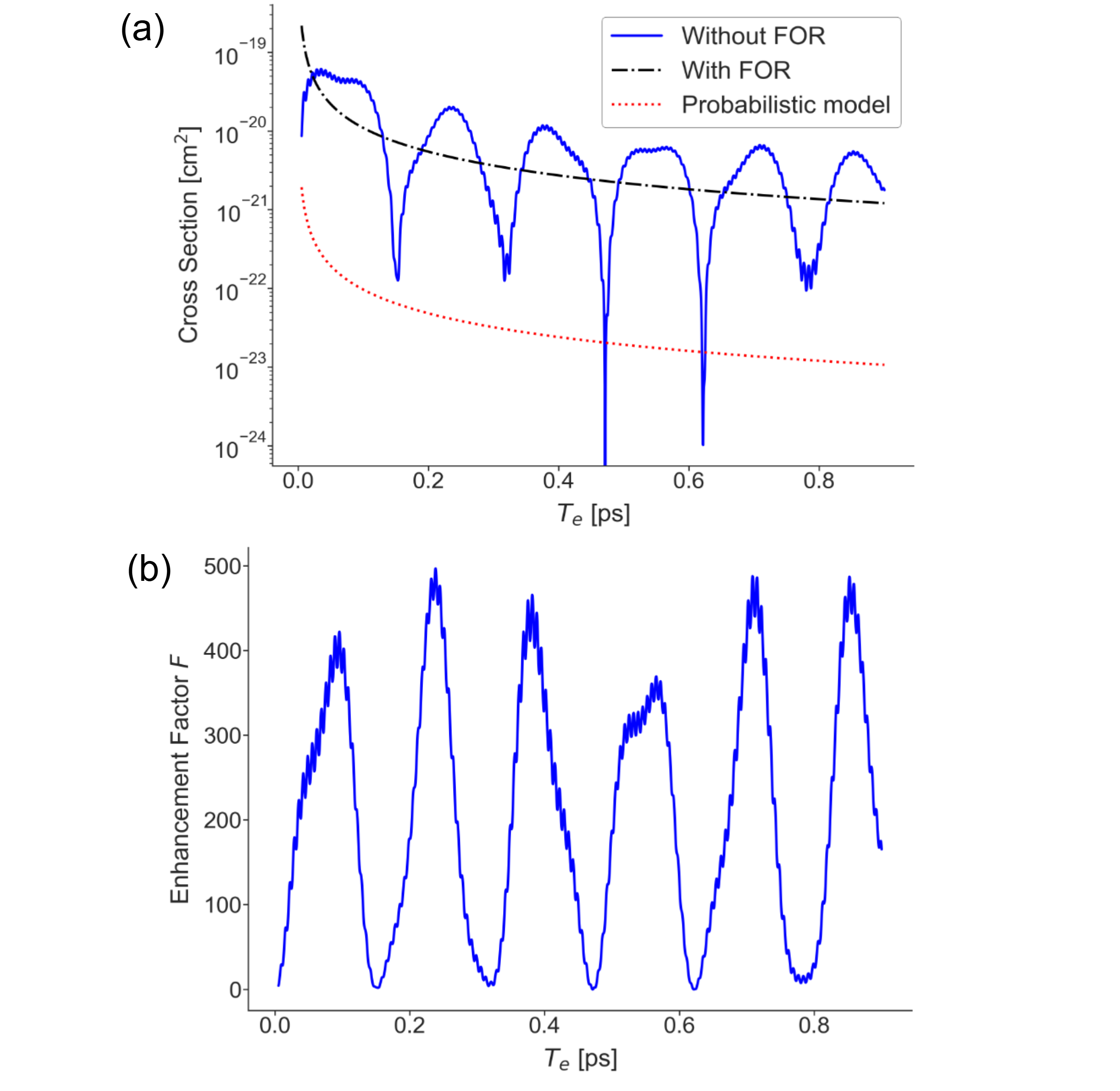}
    \caption{a) ETPA cross section as a function of the entanglement time for the three approximations considered in this work. b) shows the enhancement factor $F$ obtained without the FOR approximation as a function of the entanglement time $T_e$.}
    \label{fig:CS_complete_D0_Fixed2}
\end{figure}

In order to understand the role of the approximations on the enhancement factor $F$, we compare the values of the ETPA cross section obtained with the FOR and without the FOR approximation with the results of the probabilistic approach in Eq.~\eqref{eq:Fei_model}. Figure~\ref{fig:CS_complete_D0_Fixed2}(a) shows the dependence of $\sigma_e$ for the three cases. The dotted line corresponds to the results with probabilistic model (Eq.~\eqref{eq:Fei_model}), the dashed line to the results with the FOR approximation (Eq.~\eqref{eq:sigmae_FOR1}) and the solid line to the results without the FOR approximation (Eq.~\eqref{eq:sigmae_full}). It can be observed that the detailed calculations performed in this work result in ETPA cross section values much higher than the simple probabilistic model. The results show the importance of using a full spectrum of the entangled photons and the limitations of the FOR approximation. It is clear that with the FOR approximation any interference effects originating from the intermediate states are not present in the calculation of the cross section; in contrast with previous works that have oscillations due to additionally approximating the spectrum of the photon pairs to their central frequency~\cite{Fei}.

On the contrary, the calculation without the FOR approximation explicitly incorporates the interference mechanism between the distinct transition paths, even when considering the full spectrum of the entangled photon-pairs. In consequence, the value of the enhancement factor $F$ now has an explicit dependence on $T_e$ as shown in  Fig.~\ref{fig:CS_complete_D0_Fixed2}(b). For instance, at  $T_e\approx230$~fs the enhancement factor is $F\approx480$ due to constructive interference. On the other hand, for $T_e\approx 480$~fs, we observe the effect of entangled two-photon induced transparency, where $F<10^{-3}$. 

To estimate an experimental signal from the results in Fig.~\ref{fig:CS_complete_D0_Fixed2}, we will consider the first maximum in the enhancement factor, $T_e\approx 96$~fs with the value of $\sigma_e\approx 4.3\times10^{-20}$~cm$^2$. For a type-II SPDC process, this entanglement time corresponds to a $l=0.48$~mm crystal. The bandwidth of the photon pairs in this case is $8.1$~THz ($\approx 18$~nm)\cite{YanhuaShih_2003}. Comparing this value with the detunings $\Omega_{eg}$ in Tab.~\ref{tab:results2}, it is possible to ascertain that there is no competition with one-photon transitions when designing the experiment. To further assess the feasibility of measuring an ETPA signal, it is necessary to calculate the rate of the ETPA process. 
 Considering the number of atoms interacting with the flux of photon pairs, the total rate, $\mathcal{R}_{ETPA}$,  is 
\begin{equation}
    \mathcal{R}_{ETPA} = NV\sigma_e\phi ,
\end{equation}
where $V$ is the interaction volume of the entangled-photon pairs with the sample, and $N$ is the atomic density of atoms in the vapor cell. The interaction volume is determined by the Rayleigh range of the down-converted photons, $z_R$, and their entanglement area $A_0$. Therefore,
\begin{equation}
    \mathcal{R}_{ETPA} = 2z_RA_0N\sigma_e\phi.
    \label{eq:rate_est}
\end{equation}

\begin{figure}[ht]
    \centering
    \includegraphics[width=1.0\linewidth]{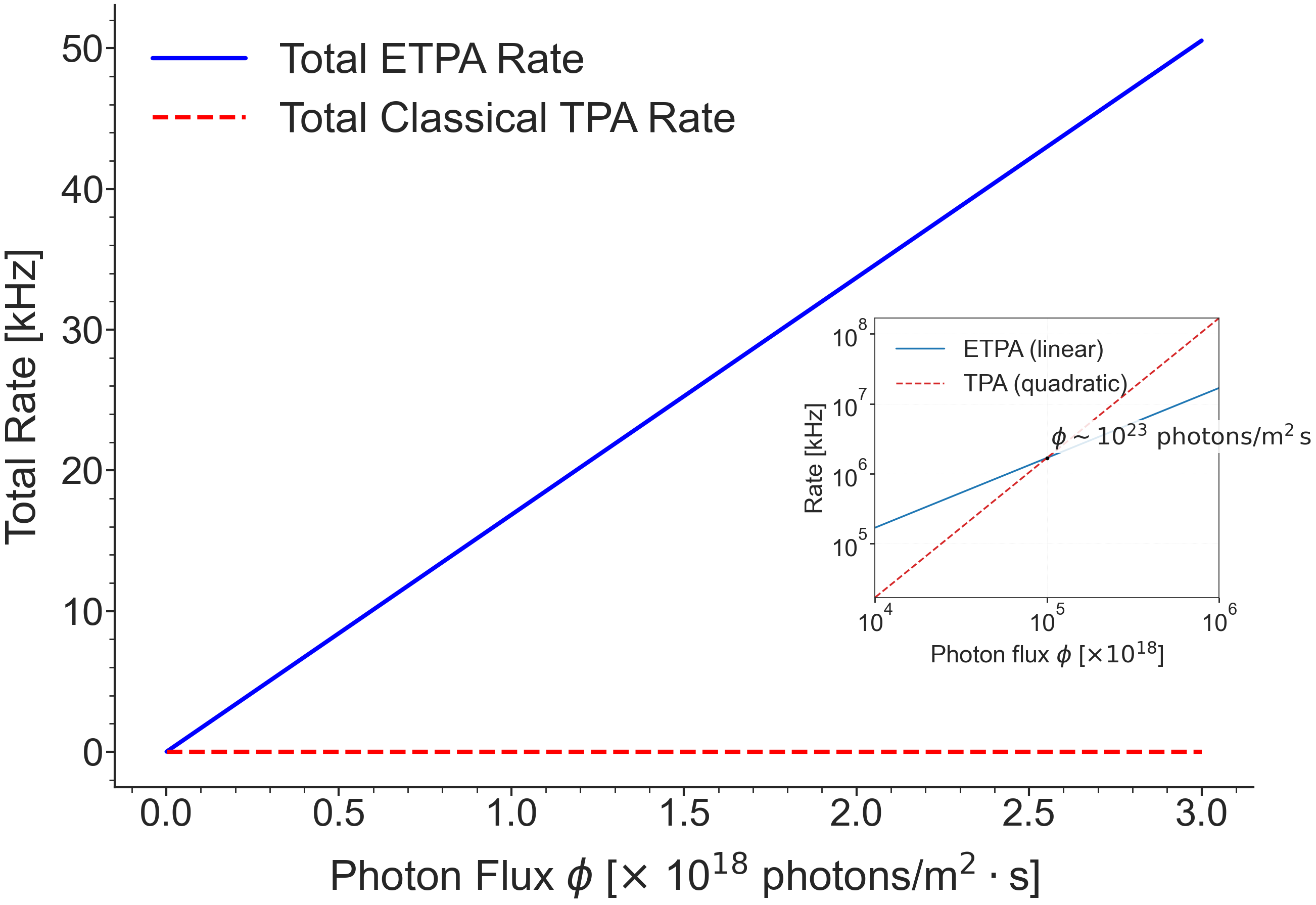}
    \caption{Total rate of the TPA and ETPA processes as a function of the photon flux. The ETPA rate is computed from Eq.~\eqref{eq:rate_est}, while the TPA rate is computed from the numbers reported in \cite{CaracasNunez:23}. For this calculation, we assumed an entanglement area of $A_0=1.8\times 10^{-8}$~m$^2$, an entanglement time $T_e=95.44$~fs, and an atomic density of $10^{19}$~atoms per m$^3$. The inset shows the same plot with values around the photon flux required to have the same rates for the TPA and ETPA processes in logarithmic scale.}
    \label{fig:Rate_comparison}
\end{figure}

For our calculation, a  $411$~nm laser is employed to pump a BBO crystal. The power of the laser is estimated to be on the order of $2$~W leading to a photon-pair flux of $\phi = 3.2\times 10^{15}$~photons$/($m$^2\cdot$s)~\cite{YanhuaShih_2003}. The photon pairs can be focused to a waist of $75$~$\mu$m which results in an entanglement area of $A_0=1.8\times 10^{-8}$~m$^2$ with a Rayleigh range of $z_R=0.01$~m. The photon pairs can interact with a hot vapor cell with an atomic density on the order of $N\approx 10^{19}$~atoms$/$m$^3$. Replacing these values in Eq.~\eqref{eq:rate_est}, it is possible to estimate an ETPA rate of $\mathcal{R}_{ETPA} \approx 50$~Hz. This signal can be significantly improved by using narrow-band pulsed lasers for producing higher fluxes of entangled-photon pairs. Figure \ref{fig:Rate_comparison} shows the total rate of the ETPA process, solid line, as a function of the flux of photon pairs. The results are compared with the rate for the classical TPA regime, dashed line, showing the advantage of exciting the transition with entangled light at this photon flux regime. 

\section{\label{sec:conclusion} Conclusion}

This work advances the theoretical understanding of the entangled two-photon absorption process by providing a detailed calculation of the ETPA cross-section for the $6^2S_{1/2}\to 8^2S_{1/2}$ transition in Cs atoms. The results account for the joint-spectrum of a type-II SPDC source of photon-pairs and for the atomic structure of cesium. An important aspect of our analysis was the comparison between the previous results using the probabilistic model~\cite{Fei}, and the values with and without the FOR approximation computed in this work. While the FOR approximation simplifies the analytical expression for the ETPA cross section, recovering the earlier probabilistic model with an enhancement factor, it averages out the individual contributions from intermediate states. The results calculated without the FOR approximation allowed us to observe oscillatory behavior in the cross section as a function of entanglement time and to identify interference effects that can lead to additional enhancement beyond the standard $F = 1$ factor. The maximum enhancement factor for the conditions given in this work is found to be close to $F\approx 500$. This discrepancy highlights the significant role of the joint spectrum of the photon pairs and the  constructive interference among different transition pathways, which only becomes apparent when the full expression of the ETPA cross-section (Eq.~\eqref{eq:brrr2}) is evaluated. These results are consistent with recent publications that suggest that making different approximations, like the spatial characteristics of the photon-pairs, taking into account finite time effects, different characteristics of the pump beam, the restriction to a single quantum path-way, among others, can change the calculated value of $\sigma_e$ in few orders of magnitudes~\cite{drago2022aspects,rodriguez2025perturbation,schlawin2024two, raymer_how_2021}. Furthermore, unifying these approaches could lead to additional enhancement factors.

Our estimation of the ETPA rate under experimentally realistic conditions provides a reference for optimizing experimental setups towards eventual observation of ETPA in atomic systems. These experiments can provide a deeper understanding of the ETPA phenomena due to their well-defined energy levels and can provide a platform for proof of principle experiments in e.g., virtual state spectroscopy~\cite{dayan2007theory, Leon2013role,Burdick2018}.

\begin{acknowledgments}

\textbf{Funding} Vice Presidency of Research and Creation at Universidad de los Andes (INV-2025-220-3489), and Faculty of Science, Universidad de los Andes (INV-2025-213-3468).\\

\textbf{Acknowledgments}. The authors thank the members of the Experimental Quantum Optics Group, Department of Physics, Universidad de los Andes, Colombia, for useful discussions regarding the results presented in this paper. The authors would like to thank the Vice Presidency of Research and Creation at Universidad de los Andes for its financial support.\\

\textbf{Disclosures}. The authors declare no conflicts of interest.\\

\textbf{Data availability}. Data underlying the results presented in this paper are not publicly available at this time but may be obtained from the corresponding author upon reasonable request.

\end{acknowledgments}

\onecolumngrid
\appendix

\section{Quantization of the Electromagnetic Field}
\label{sec:AppendA}

Consider an electromagnetic field propagating in vacuum. This field can be described using the vector potential $\mathbf{A}(\mathbf{r},t)$, which satisfies the wave equation~\cite{Gerry_Knight_2005}
\begin{equation}
    \nabla^2\mathbf{A} - \frac{1}{c^2}\frac{\partial^2 \mathbf{A}}{\partial t^2} = 0,
    \label{eq:wave_equation}
\end{equation}
and the Coulomb gauge condition
\begin{equation}
    \nabla\cdot\mathbf{A}(\mathbf{r},t) = 0.
    \label{eq:coulomb_gauge}
\end{equation}

The vector potential can be expressed as a superposition of plane waves
\begin{equation}
    \mathbf{A(\mathbf{r},t)} = \sum_j \mathbf{e}_j\left[ A_j(t)e^{i\mathbf{k}_j\cdot \mathbf{r}} + A_j^*(t)e^{-i\mathbf{k}_j\cdot \mathbf{r}} \right],
    \label{eq:Potential_A}
\end{equation}
where, $A_j(t)$ is the complex amplitude, $\mathbf{e}_j$ is the polarization vector, and $\mathbf{k}_j$ is the wave vector of the $j$th mode. From here onward, we will assume that all modes have the same polarization, i.e, $\mathbf{e}_j = \mathbf{e}$. From Eqs.~\eqref{eq:wave_equation} and \eqref{eq:coulomb_gauge}, the amplitude $A_j(t)$ satisfies 
\begin{equation}
    \frac{\mathrm{d}^2 A_j}{\mathrm{d}t^2} + \omega_j^2 A_j = 0,
\end{equation}
where $\omega_j = ck_j$ is the mode frequency. The solution to this equation is
\begin{equation}
	A_j(t) = A_je^{-i\omega_jt}.
\end{equation}

The electric field follows from the vector potential as
\begin{equation}
    \mathbf{E}(\mathbf{r},t) = -\frac{\partial \mathbf{A}}{\partial t}.
    \label{eq:cl_electricf}
\end{equation}

Substituting Eq.~\eqref{eq:Potential_A}, we obtain
\begin{equation}
    \mathbf{E}(\mathbf{r},t) = i\sum_j\omega_j\mathbf{e}\left[ A_je^{i(\mathbf{k}_j\cdot \mathbf{r}-\omega_jt)} - A_j^*e^{-i(\mathbf{k}_j\cdot \mathbf{r}-\omega_jt)} \right].
    \label{eq:ElectricField}
\end{equation}

Assuming that the electromagnetic field propagates through a region in space of volume $V$, the Hamiltonian of the electromagnetic field is given by~\cite{Gerry_Knight_2005}
\begin{equation}
    H = 2\epsilon_0 V\sum_j\omega_j^2A_j^*A_j,
    \label{eq:Em_energy2}
\end{equation}
where $\epsilon_0$ is the vacuum permittivity. This expression reveals that each mode of the electromagnetic field behaves as a harmonic oscillator, with the summation accounting for the energy contributions of all individual modes.\\

To quantize the field, we introduce the canonical variables $p_j$ and $q_j$ via~\cite{Gerry_Knight_2005}:
\begin{align}
    A_j &= \frac{1}{2 \omega_j (\epsilon_0 V)^{1/2}} (\omega_j q_j + i p_j),
    \label{eq:A_j} \\
    A_j^* &= \frac{1}{2 \omega_j (\epsilon_0 V)^{1/2}} (\omega_j q_j - i p_j).
    \label{eq:A_j*}
\end{align}

Substituting into Eq.~\eqref{eq:Em_energy2}, the Hamiltonian reduces to
\begin{equation}
    H = \frac{1}{2}\sum_j(p_j^2 + \omega_j^2q_j^2).
    \label{eq:AO_energy}
\end{equation}

The next step is to transform the canonical variables $q$ and $p$ into operators $\hat{q}$ and $\hat{p}$ which satisfy the canonical commutation relations
\begin{align}
    \left[ \hat{q}_i, \hat{q}_j \right] &= \left[ \hat{p}_i, \hat{p}_j \right] = 0, \\
    \left[ \hat{q}_i, \hat{p}_j \right] &= i\hslash\delta_{ij}.
\end{align}

The annihilation and creation operators are then defined as~\cite{Gerry_Knight_2005}
\begin{align}
    \hat{a}_j &= \frac{1}{(2 \hbar \omega_j)^{1/2}} (\omega_j \hat{q}_j + i \hat{p}_j),
    \label{eq:a_operator} \\
    \hat{a}_j^{\dagger} &= \frac{1}{(2 \hbar \omega_j)^{1/2}} (\omega_j \hat{q}_j - i \hat{p}_j),
    \label{eq:a+_operator}
\end{align}
with commutation relations
\begin{align}
    \left[ \hat{a}_l, \hat{a}_m \right] &= \left[ \hat{a}_l^{\dagger}, \hat{a}_m^{\dagger} \right] = 0, \notag \\
    \left[ \hat{a}_l, \hat{a}_m^{\dagger} \right] &= \delta_{lm}.
    \label{eq:canonical_commutation}
\end{align}

Substituting Eqs.~\eqref{eq:a_operator} and \eqref{eq:a+_operator} into the Eqs.~\eqref{eq:A_j} and \eqref{eq:A_j*}, the amplitudes of the vector potential can be written as operators in terms of the annihilation and creation operators:
\begin{equation}
    \hat{A}_j = \left( \frac{\hslash}{2\omega_j\epsilon_0V} \right)^{1/2}\hat{a}_j,
\end{equation}
and thus, the quantized vector potential takes the form
\begin{equation}
    \hat{\mathbf{A}}(\mathbf{r},t) = \sum_j\left( \frac{\hslash}{2\omega_j\epsilon_0V} \right)^{1/2}\mathbf{e}\left[ \hat{a}_je^{i(\mathbf{k}_j\cdot \mathbf{r}-\omega_jt)} + \hat{a}_j^{\dagger}e^{-i(\mathbf{k}_j\cdot \mathbf{r}-\omega_jt)} \right].
\end{equation}

Following Eq.~\eqref{eq:ElectricField}, the electric field operator can be written as
\begin{equation}
    \hat{\mathbf{E}}(\mathbf{r},t) = \hat{\mathbf{E}}^{(+)}(\mathbf{r},t) + \hat{\mathbf{E}}^{(-)}(\mathbf{r},t),
\end{equation}
where
\begin{equation}
    \hat{\mathbf{E}}^{(+)}(\mathbf{r},t) = i\sum_j\left( \frac{\hslash\omega_j}{2\epsilon_0V} \right)^{1/2}\hat{a}_je^{i(\mathbf{k}_j\cdot \mathbf{r}-\omega_jt)}\mathbf{e},
    \label{eq:quantized_E}
\end{equation}
and
\begin{equation}
    \hat{\mathbf{E}}^{(-)}(\mathbf{r},t) = \left[ \hat{\mathbf{E}}^{(+)}(\mathbf{r},t) \right]^{\dagger}.
\end{equation}

The plane waves of Eq.~\eqref{eq:quantized_E} are assumed to propagate along the $z$ axis through an empty optical cavity of length $\tilde{l}$ in the limit $\tilde{l}\to \infty$. The mode spectrum becomes continuum as $\tilde{l}\to \infty$. Therefore, the eigen-modes of the field are separated by \cite{Blow90}
\begin{equation}
    \Delta k = \frac{2\pi}{\tilde{l}},
\end{equation}
and frequency
\begin{equation}
    \Delta \omega = \frac{2\pi c}{\tilde{l}}.
\end{equation}

In this limit, $\Delta\omega\to 0$ and the annihilation operators in Eq.~\eqref{eq:quantized_E} are transformed into continuous-mode operators following the convention used in~\cite{Blow90}:
\begin{equation}
    \hat{a}_j \to \sqrt{\frac{\Delta\omega}{2\pi}}\hat{a}(\omega),
    \label{eq:a_to_aw}
\end{equation}
and the commutation relation in Eq.~\eqref{eq:canonical_commutation} becomes
\begin{equation}
    \left[ \hat{a}(\omega), \hat{a}^{\dagger}(\omega') \right] = 2\pi\delta(\omega - \omega'),
    \label{eq:commute_rel}
\end{equation}
and sums over discrete modes are converted into integrals over the continuous frequency by 
\begin{equation}
    \sum_j \to \frac{1}{\Delta\omega} \int_{0}^{\infty}\mathrm{d}\omega.
\end{equation}

The electric field operator in Eq.~\eqref{eq:quantized_E} is transformed into
\begin{equation}
    \hat{\mathbf{E}}^{(+)}(z,t) = i\int_{0}^{\infty}\frac{\mathrm{d}\omega}{2\pi}\left( \frac{\hslash\omega}{2\epsilon_0cA_0} \right)^{1/2}\hat{a}(\omega)e^{i\left[k(\omega)z-\omega t\right]}\mathbf{e},
    \label{eq:quantized_EM4A}
\end{equation}
where we assumed that $\mathbf{k}\cdot\mathbf{r}=kz$ and $A_0=\tilde{l}^2$. For mathematical convenience, we extend the integration limits to include negative frequencies, even though they have no physical significance. Since the frequency spectra used in the calculations contribute minimally in this region, the resulting error is negligible. Therefore,
\begin{equation}
    \hat{\mathbf{E}}^{(+)}(z,t) = i\int_{-\infty}^{\infty}\frac{\mathrm{d}\omega}{2\pi}\left( \frac{\hslash\omega}{2\epsilon_0cA_0} \right)^{1/2}\hat{a}(\omega)e^{i\left[k(\omega)z-\omega t\right]}\mathbf{e}.
    \label{eq:quantized_EM5}
\end{equation}

\section{Calculation of the Atomic Correlation Function}
\label{append:C}

In this Appendix, it is demonstrated how to get the explicit form of the atomic correlation function $C_A(t_1,t_2,t_3,t_4)$ in Eq.~\eqref{eq:atomcorrelation} starting from Eq.~\eqref{eq:CM_1}:
\begin{equation}
    C_A(t_1,t_2,t_3,t_4) =\trace_A\left[ \hat{d}^{(-)}(t_4)\hat{d}^{(-)}(t_3)\hat{\rho}_A(0)\hat{d}^{(+)}(t_1)\hat{d}^{(+)}(t_2)\ket{f}\bra{f} \right ].
    \label{eq:CM_1A}
\end{equation}

It is noted first, taking $\hat{\rho}_A(0) = \ket{g}\bra{g}$, that
\begin{align}
    C_A(t_1,t_2,t_3,t_4) &= \trace_A\left[ \hat{d}^{(-)}(t_4)\hat{d}^{(-)}(t_3)\ket{g}\bra{g}\hat{d}^{(+)}(t_1)\hat{d}^{(+)}(t_2)\ket{f}\bra{f} \right], \\
    &= \bra{f}\hat{d}^{(-)}(t_4)\hat{d}^{(-)}(t_3)\ket{g}\bra{g}\hat{d}^{(+)}(t_1)\hat{d}^{(+)}(t_2)\ket{f}, \\
    &= \sum_{e'} \bra{f}\hat{d}^{(-)}(t_4)\ket{e'}\bra{e'}\hat{d}^{(-)}(t_3)\ket{g}\sum_{e} \bra{g}\hat{d}^{(+)}(t_1)\ket{e}\bra{e}\hat{d}^{(+)}(t_2)\ket{f}.
\end{align}

Using the definition of the electric dipole operator in Eq.~\eqref{eq:dipole_operator} it follows that
\begin{equation}
    C_A(t_1,t_2,t_3,t_4) = \sum_{e,e'}d_{fe'}d_{e'g}d_{ge}d_{ef}e^{i(\omega_f - \omega_{e'})t_4}e^{i(\omega_{e'} - \omega_g)t_3}e^{-i(\omega_e - \omega_g)t_1}e^{-i(\omega_f - \omega_e)t_2}.
\end{equation}

Introducing the change of variables $\tau = t_2-t_1$, $s=t_3-t_2$ and $r=t_4-t_3$, defining $\omega_{fe'} = \omega_f - \omega_{e'}$, $\omega_{fg} = \omega_f - \omega_g$ and $\omega_{eg} = \omega_e - \omega_g$, and simplifying terms it is obtained that
\begin{equation}
     C_A(t_1,t_2,t_3,t_4) = \sum_{e,e'}d_{fe'}d_{e'g}d_{ge}d_{ef}e^{i\omega_{fe'}r}e^{i\omega_{fg}s}e^{i\omega_{eg}\tau}.
\end{equation}

The Kubo theory for the line shape allows to include decoherence mechanisms by adding an imaginary term $i\gamma_{jk}$ to the transition frequencies such that $\omega_{jk}$ becomes $\omega_j-\omega_k+i\gamma_{jk}$~\cite{Raymer2021, kubo1969stochastic}. Taking into account these dephasing rates, the atomic correlation function is
\begin{equation}
     C_A(t_1,t_2,t_3,t_4) = \sum_{e,e'}d_{fe'}d_{e'g}d_{ge}d_{ef}e^{-(\gamma_{fe'} - i\omega_{fe'})r}e^{-(\gamma_{fg} - i\omega_{fg})s}e^{-(\gamma_{eg} - i\omega_{eg})\tau}.
     \label{eq:CM_A}
\end{equation}

\section{Calculation of the Field Correlation Function}
\label{append:B}

\subsection{Quantum state of photon pairs produced by SPDC}

In this appendix, the biphoton state, $\ket{\Psi}$, and the field correlation function, $C_F(t_1,t_2,t_3,t_4)$, of the down-converted photons is presented, following the references~\cite{grice1997spectral, YanhuaShih_2003}. The quantum state of the entangled-photon pairs produced in the SPDC process can be written as
\begin{equation}
    \ket{\Psi} = \text{exp}\left[ \frac{1}{i\hslash}\int_{t_0}^{t}dt'\hat{H}_{SPDC}(t') \right]\ket{0}, 
    \label{eq:biphoton_state}
\end{equation}
where we assume that in the initial time $t_0$, the state of the photon-pairs corresponds to the vacuum state $\ket{0}$. Using first order perturbation theory, the biphoton state is
\begin{equation}
    \ket{\tilde{\Psi}} =\ket{0} +\frac{1}{i\hslash}\int_{t_0}^{t}dt'\hat{H}_{SPDC}(t') \ket{0},
    \label{eq:biphoton_state2}
\end{equation}
with $\hat{H}_{SPDC}(t)$ as the interaction Hamiltonian associated to the SPDC process, which is given by~\cite{YanhuaShih_2003,grice1997spectral}.
\begin{equation}
    \hat{H}_{SPDC}(t)=\epsilon_0\int_{V}d^{3}\mathbf{r} \chi^{(2)}\hat{E}_p^{(+)}(\mathbf{r},t)\hat{E}_s^{(-)}(\mathbf{r},t)\hat{E}_i^{(-)}(\mathbf{r},t) + h.c.
    \label{eq:Hamil_SPDC2}
\end{equation}

The pump light is assumed to be a classical electromagnetic wave with a certain spectral distribution $f(\omega_p)$ that propagates along the $z$ direction. Therefore,
\begin{equation}
    \hat{E}_p^{(+)}(z,t) \approx E_p(z,t) = \int_{-\infty}^{\infty} E_pf(\omega_p)e^{i(k_pz - \omega_pt)}d\omega_p,
    \label{eq:pump_field}
\end{equation}
where $E_p$ is the amplitude of the electric field of the pump beam. Considering a SPDC process where the photons are emitted in a collinear configuration, the electric field operator for the signal and idler photons is given by
\begin{equation}
    \hat{E}_{s,i}^{(-)}(z,t) = -i\int_{-\infty}^{\infty}\frac{\mathrm{d}\omega_{s,i}}{2\pi}\left( \frac{\hslash\omega_{s,i}}{2\epsilon_0cA_0} \right)^{1/2}\hat{a}(\omega_{s,i})e^{-i(k_{s,i}z-\omega_{s,i}t)}.
    \label{eq:quantized_EM5B}
\end{equation}

For a thin non-linear crystal, it is valid to approximate the Hamiltonian to its longitudinal part \cite{YanhuaShih_2003}, therefore,
\begin{align}
  \frac{1}{i\hbar} \int_{-\infty}^{\infty} \hat{H}_{SPDC}(t)dt&=-\frac{\epsilon_0\chi^{(2)}}{i\hbar}\int_{-\infty}^{\infty} E_pf(\omega_p)d\omega_p\int_{-\infty}^{\infty}\frac{\mathrm{d}\omega_s}{2\pi}\left( \frac{\hslash\omega_s}{2\epsilon_0cA_0} \right)^{1/2}\hat{a}(\omega_s)\int_{-\infty}^{\infty}\frac{\mathrm{d}\omega_i}{2\pi}\left( \frac{\hslash\omega_i}{2\epsilon_0cA_0} \right)^{1/2}\hat{a}(\omega_i)\notag\\
    &\times\int _{-l/2}^{l/2}dze^{i(k_p-k_s-k_i)z}\int_{-\infty}^{\infty}dte^{-i(\omega_p-\omega_s-\omega_i)t} + h.c.
    \label{eq:Hamil_SPDC2}
\end{align}

The spatial integral is 
\begin{equation}
    \frac{1}{l}\int_{-l/2}^{l/2}\mathrm{d}ze^{-i(k_s+k_i-k_p)z} = \frac{\text{sin}[(k_s+k_i-k_p)l/2]}{(k_s+k_i-k_p)l/2}.
    \label{eq:sinc_A}
\end{equation}
and the time integral results in 
\begin{equation}
    \int_{-\infty}^{\infty}\mathrm{d}te^{-i(\omega_p-\omega_s-\omega_i)t} = 2\pi\delta(\omega_p-\omega_s-\omega_i).
    \label{eq:energy_conserv}
\end{equation}

This last expression is the phase-matching condition for energy conservation, which implies that 
\begin{equation}
    \omega_s + \omega_i = \omega_p.
\end{equation}

In the frequency domain, it is valid to define a central frequency for the entangled photon-pairs, $\omega_0$, such that 
\begin{equation}
    \left\{\begin{matrix}
        \omega_s&= \omega_0 + \Omega_s, \\ 
	\omega_i&= \omega_0 - \Omega_i, \\ 
	\omega_p&= 2\omega_0 + \Omega_p, 
    \end{matrix}\right.
    \label{eq:detunings}
\end{equation}
where $\Omega_s, \Omega_i, \Omega_p$ are detunings of the \textit{signal}, \textit{idler} and pump frequencies around their central frequencies. Considering $\Omega_{s,i}\ll\omega_0$, the following approximation can be applied:
\begin{equation}
    \left( \frac{\hslash\omega_{s,i}}{2\epsilon_0cA_0} \right)^{1/2} \approx \left( \frac{\hslash\omega_0}{2\epsilon_0cA_0} \right)^{1/2}\left( 1 \pm \alpha_{s,i} \right).
\end{equation}

Here $\alpha_{s,i}=\Omega_{s,i}/\omega_0$ is an error factor associated with how small $\Omega_{s,i}$ is with respect to $\omega_0$. From here on, $\alpha_{s,i}\ll 1$ will be assumed. Therefore, the following constant can be defined
\begin{equation}
    L_0 = \left( \frac{\hslash\omega_0}{2\epsilon_0cA_0} \right)^{1/2},
    \label{eq:L0}
\end{equation}
such that

\begin{equation}
    \small
    \frac{1}{i\hbar} \int_{-\infty}^{\infty} \hat{H}_{SPDC}(t)dt=-\frac{\epsilon_0\chi^{(2)}l}{i\hbar 2\pi}L_0^2\int_{-\infty}^{\infty}\mathrm{d}\omega_s\hat{a}(\omega_s)\int_{-\infty}^{\infty}\mathrm{d}\omega_i\hat{a}(\omega_i)\int_{-\infty}^{\infty} E_pf(\omega_p)d\omega_p \frac{\text{sin}[(k_s+k_i-k_p)l/2]}{(k_s+k_i-k_p)l/2}\delta(\omega_p-\omega_s-\omega_i)+ h.c.
\end{equation}

From this expression it is possible to define the Joint-Spectral Amplitude (JSA) function of the biphoton state as
\begin{equation}
    \Phi(\omega_s, \omega_i) =C \int_{-\infty}^{\infty}\mathrm{d}\omega_pf(\omega_p)\delta(\omega_p-\omega_s-\omega_i)\frac{\text{sin}[(k_s+k_i-k_p)l/2]}{(k_s+k_i-k_p)l/2},
    \label{eq:JSA_A1}
\end{equation}
with $C$ as a normalization constant. The spectral properties of the biphoton state are contained in this function.  Thus, the state of the biphoton can be written as, 
\begin{equation}
    \ket{\tilde{\Psi}} \approx \ket{0} + K\int_{-\infty}^{\infty}\mathrm{d}\omega_s\int_{-\infty}^{\infty}\mathrm{d}\omega_i\hat{a}^{\dagger}(\omega_s)\hat{a}^{\dagger}(\omega_i)\Phi(\omega_s,\omega_i)\ket{0},
    \label{eq:biphoton_state2}
\end{equation}
with
\begin{equation}
    K = -\frac{\epsilon_0\chi^{(2)}L_0^2E_pl}{2\pi C i\hslash}.
\end{equation}

In order to normalize the state in Eq. \eqref{eq:biphoton_state2}, lets begin by defining the operator 
\begin{equation}
    \hat{I} = \int_{-\infty}^{\infty}\mathrm{d}\omega_s\int_{-\infty}^{\infty}\mathrm{d}\omega_i\hat{a}^{\dagger}(\omega_s)\hat{a}^{\dagger}(\omega_i)\Phi(\omega_s,\omega_i),
\end{equation}
so that
\begin{equation}
    \ket{\tilde{\Psi}} = \ket{0} + K\hat{I}\ket{0},
\end{equation}
and
\begin{equation}
     \braket{\tilde{\Psi}}{\tilde{\Psi}}= \braket{0}{0} + \bra{0}K^*\hat{I}^{\dagger}\ket{0} + \bra{0}K\hat{I}\ket{0} + \bra{0}|K|^2\hat{I}^{\dagger}\hat{I}\ket{0}.
\end{equation}

Noting that $\braket{0}{0} = 1$ and $\bra{0}K^*\hat{I}^{\dagger}\ket{0} = \bra{0}K\hat{I}\ket{0} = 0$, it follows that
\begin{equation}
    \braket{\tilde{\Psi}}{\tilde{\Psi}} = 1 + |K|^2\bra{0}\hat{I}^{\dagger}\hat{I}\ket{0}.
    \label{eq:norm_step1}
\end{equation}

The next step is to compute the term $\hat{I}^{\dagger}\hat{I}\ket{0}$ using the commutation relation in Eq.~\eqref{eq:commute_rel} between the operators $\hat{a}(\omega_s')$ and $\hat{a}^{\dagger}(\omega_s)$. As a result, we can see that
\begin{align}
\hat{I}^{\dagger}\hat{I}\ket{0} &= \int_{-\infty}^{\infty}\mathrm{d}\omega_s'\int_{-\infty}^{\infty}\mathrm{d}\omega_i'\int_{-\infty}^{\infty}\mathrm{d}\omega_s\int_{-\infty}^{\infty}\mathrm{d}\omega_i\hat{a}(\omega_i')\left(\hat{a}(\omega_s')\hat{a}^{\dagger}(\omega_s)\right)\hat{a}^{\dagger}(\omega_i)\Phi^*(\omega_s',\omega_i')\Phi(\omega_s,\omega_i)\ket{0}\\
     &= \int_{-\infty}^{\infty}\mathrm{d}\omega_s'\int_{-\infty}^{\infty}\mathrm{d}\omega_i'\int_{-\infty}^{\infty}\mathrm{d}\omega_s\int_{-\infty}^{\infty}\mathrm{d}\omega_i 2\pi\delta(\omega_s'-\omega_s)\left(\hat{a}(\omega_i')\hat{a}^{\dagger}(\omega_i)\right)\Phi^*(\omega_s',\omega_i')\Phi(\omega_s,\omega_i)\ket{0} \notag \\
    &+ \int_{-\infty}^{\infty}\mathrm{d}\omega_s'\int_{-\infty}^{\infty}\mathrm{d}\omega_i'\int_{-\infty}^{\infty}\mathrm{d}\omega_s\int_{-\infty}^{\infty}\mathrm{d}\omega_i\hat{a}(\omega_i')\hat{a}^{\dagger}(\omega_s)\left(\hat{a}(\omega_s')\hat{a}^{\dagger}(\omega_i)\right)\Phi^*(\omega_s',\omega_i')\Phi(\omega_s,\omega_i)\ket{0}.
\end{align}

In the first term, we use the commutation relation~\eqref{eq:commute_rel} for $\hat{a}(\omega_i')\hat{a}^{\dagger}(\omega_i)$, while in the second term we apply it to the product $\hat{a}(\omega_s')\hat{a}^{\dagger}(\omega_i)$:
\begin{align}
    \hat{I}^{\dagger}\hat{I}\ket{0} &= 2\pi\int_{-\infty}^{\infty}\mathrm{d}\omega_s'\int_{-\infty}^{\infty}\mathrm{d}\omega_i'\int_{-\infty}^{\infty}\mathrm{d}\omega_s\int_{-\infty}^{\infty}\mathrm{d}\omega_i\delta(\omega_s'-\omega_s)\left[ 2\pi\delta(\omega_i' - \omega_i) + \hat{a}^{\dagger}(\omega_i)\hat{a}(\omega_i') \right]\Phi^*(\omega_s',\omega_i')\Phi(\omega_s,\omega_i)\ket{0} \notag \\
    &+ \int_{-\infty}^{\infty}\mathrm{d}\omega_s'\int_{-\infty}^{\infty}\mathrm{d}\omega_i'\int_{-\infty}^{\infty}\mathrm{d}\omega_s\int_{-\infty}^{\infty}\mathrm{d}\omega_i\hat{a}(\omega_i')\hat{a}^{\dagger}(\omega_s)\left[ 2\pi\delta(\omega_s' - \omega_i) + \hat{a}^{\dagger}(\omega_i)\hat{a}(\omega_s') \right]\Phi^*(\omega_s',\omega_i')\Phi(\omega_s,\omega_i)\ket{0}.
\end{align}

The second term in the first line is zero because $\hat{a}(\omega_i')\ket{0} = 0$. The same follows for $\hat{a}(\omega_s')\ket{0}$ in the second line. Therefore,
the first term reduces to $(2\pi)^2\int_{-\infty}^{\infty}\mathrm{d}\omega_s\int_{-\infty}^{\infty}\mathrm{d}\omega_i\left| 
\Phi(\omega_s,\omega_i) \right|^2\ket{0}$. For the second term we apply again the commutation relation~\eqref{eq:commute_rel} and discard the $\hat{a}(\omega_i')\ket{0}$ term to see that
\begin{align}
    \hat{I}^{\dagger}\hat{I}\ket{0} &= (2\pi)^2\int_{-\infty}^{\infty}\mathrm{d}\omega_s\int_{-\infty}^{\infty}\mathrm{d}\omega_i\left|\Phi(\omega_s,\omega_i) \right|^2\ket{0} \notag \\
    &+ (2\pi)^2\int_{-\infty}^{\infty}\mathrm{d}\omega_s'\int_{-\infty}^{\infty}\mathrm{d}\omega_i'\int_{-\infty}^{\infty}\mathrm{d}\omega_s\int_{-\infty}^{\infty}\mathrm{d}\omega_i\delta(\omega_s' - \omega_i)\delta(\omega_i' - \omega_s)\Phi^*(\omega_s',\omega_i')\Phi(\omega_s,\omega_i)\ket{0}.
\end{align}

Assuming that the JSA function is normalized, i.e, $\int_{-\infty}^{\infty}\mathrm{d}\omega_s\int_{-\infty}^{\infty}\mathrm{d}\omega_i\left|\Phi(\omega_s,\omega_i) \right|^2 = 1$, we get finally that
\begin{align}
   \braket{\tilde{\Psi}}{\tilde{\Psi}} &= 1 + |K|^2\bra{0}\hat{I}^{\dagger}\hat{I}\ket{0}\\
   &=1+ (2\pi|K|)^2\left[ 1 + \int_{-\infty}^{\infty}\mathrm{d}\omega_s\int_{-\infty}^{\infty}\mathrm{d}\omega_i\Phi^*(\omega_i,\omega_s)\Phi(\omega_s,\omega_i) \right]\\
   &= 1 + |\beta|^2.
\end{align}

The normalized biphoton state is then obtained by dividing Eq.~\eqref{eq:biphoton_state2} by $\sqrt{1+|\beta|^2}$:
\begin{equation}
	\ket{\Psi} = \frac{1}{\sqrt{1+|\beta|^2}}\ket{0} + \frac{\beta}{\sqrt{1+|\beta|^2}}\int_{-\infty}^{\infty}\mathrm{d}\omega_s\int_{-\infty}^{\infty}\mathrm{d}\omega_i\hat{a}^{\dagger}(\omega_s)\hat{a}^{\dagger}(\omega_i)\Phi(\omega_s,\omega_i)\ket{0}.
\end{equation}

Defining the constant
\begin{equation}
	\varepsilon = \frac{\beta}{\sqrt{1+|\beta|^2}},
\end{equation}
we can finally write the normalized biphoton state as
\begin{equation}
    \ket{\Psi} =\sqrt{1-|\varepsilon|^2}\ket{0} + \varepsilon\int_{-\infty}^{\infty}\mathrm{d}\omega_s\int_{-\infty}^{\infty}\mathrm{d}\omega_i\hat{a}^{\dagger}(\omega_s)\hat{a}^{\dagger}(\omega_i)\Phi(\omega_s,\omega_i)\ket{0}.
\end{equation}

In the SPDC process, the first term represents the fraction of the amplitude of the pump field that did not interact with the crystal. As a consequence, for our calculations we can write the biphoton state as
\begin{equation}
    \ket{\Psi} = \varepsilon\int_{-\infty}^{\infty}\mathrm{d}\omega_s\int_{-\infty}^{\infty}\mathrm{d}\omega_i\hat{a}^{\dagger}(\omega_s)\hat{a}^{\dagger}(\omega_i)\Phi(\omega_s,\omega_i)\ket{0}.
    \label{eq:Normalized_bs}
\end{equation}

Given that $\varepsilon$ is the probability of producing down-converted photons in the SPDC process, the number of photon-pairs is then proportional to $|\varepsilon|^2$. 

\subsection{Computation of the JSA Function}

To compute the JSA function $\Phi(\omega_s,\omega_i)$ in Eq.~\eqref{eq:JSA_A1} it is necessary to write the wave numbers in terms of the frequencies $\omega_s$, $\omega_i$ and $\omega_p$. Due to the dispersion of the photon pairs in the crystal, the wavenumbers are expanded in a Taylor series around the central frequencies, $\omega_0$ and $2\omega_0$ (Eq.~\eqref{eq:detunings}). If the entangled-photon pairs are generated via a type-II SPDC process, the expansion up to first order is~\cite{YanhuaShih_2003}:
\begin{align}
    k_s &= k_s(\omega_0) + k_s'\Omega_s, \\ 
    k_i &= k_i(\omega_0) - k_i'\Omega_i, \\ 
    k_p &= k_p(2\omega_0) + k_p'\Omega_p, 
\end{align}
where the $k_m'$ are the group velocities of the signal, idler and pump photons. Assuming perfect momentum phase-matching at the central frequency, $k_s(\omega_0) + k_i(\omega_0) = k_p(2\omega_0)$, and considering the Eq. \eqref{eq:detunings}, the JSA function takes the form
\begin{equation}
\Phi(\omega_s,\omega_i) = \Phi(\Omega_s+\omega_0,\omega_0-\Omega_i) = C\int_{-\infty}^{\infty}\mathrm{d}\Omega_pf(\Omega_p)\delta(\Omega_p - \Omega_s + \Omega_i)\frac{\text{sin}\left[ \left( k_s'\Omega_s-k_i'\Omega_i-k_p'\Omega_p \right)l/2 \right]}{\left( k_s'\Omega_s-k_i'\Omega_i-k_p'\Omega_p \right)l/2}.
	\label{eq:JSA2}
\end{equation}

Integrating over $\Omega_p$ it is obtained that
\begin{equation}
	\Phi(\Omega_s+\omega_0,\omega_0-\Omega_i) = Cf(\Omega_s - \Omega_i)\frac{\text{sin}\left[ \left( (k_s'-k_p')\Omega_s+(k_p'-k_i')\Omega_i\right)l/2 \right]}{\left( (k_s'-k_p')\Omega_s+(k_p'-k_i')\Omega_i\right)l/2}.
	\label{eq:JSA4}
\end{equation}

For computing the ETPA cross-section, it will become useful to write this JSA function as
\begin{equation}
    \Phi(\Omega_s+\omega_0,\omega_0-\Omega_i) = f(\Omega_s-\Omega_i)\varphi(\Omega_s,\Omega_i),
    \label{eq:type-II-JSA1}
\end{equation}
with
\begin{equation}
    \varphi(\Omega_s,\Omega_i) = C\frac{\text{sin}\left[ \left( (k_s'-k_p')\Omega_s+(k_p'-k_i')\Omega_i\right)l/2 \right]}{\left( (k_s'-k_p')\Omega_s+(k_p'-k_i')\Omega_i\right)l/2},
    \label{eq:varphi}
\end{equation}

In this paper, we will restrict to the case where the entangled-photon pairs are generated with a monochromatic pump beam, such that 
\begin{equation}
	|f(\omega_p)|^2 =\delta(\omega_p -2\omega_0) = |f(\Omega_s-\Omega_i)|^2= \delta(\Omega_p)= \delta(\Omega_s-\Omega_i),
    \label{eq:delta}
\end{equation}
and the normalization constant $C$ can be computed as follows:
\begin{equation}
	\frac{1}{C^2} = \int_{-\infty}^{\infty}\mathrm{d}\Omega_i\int_{-\infty}^{\infty}\mathrm{d}\Omega_s|\Phi(\Omega_s+\omega_0,\omega_0-\Omega_i)|^2.
\end{equation}

Performing the integration over $\Omega_s$, it can be seen that
\begin{equation}
	\frac{1}{C^2} = \int_{-\infty}^{\infty}\text{sinc}^2[(k_s'-k_i')\Omega_i l/2]\mathrm{d}\Omega_i.
\end{equation}
where the $\text{sinc}(x)=\text{sin}(x)/x$. Therefore,
\begin{equation}
    C = \sqrt{\frac{T_e}{2\pi}}.
\end{equation}

The entanglement time for the photons produced in a type-II SPDC crystal is defined as $T_e= l(k_s' - k_i')$.

\subsection{Calculation of the field correlation function}
Once the biphoton state $\ket{\Psi}$ is computed, the field correlation function can be calculated according to Eq.~\eqref{eq:CF_1}:

\begin{equation}
     C_F(t_1,t_2,t_3,t_4) = \bra{\Psi}\hat{E}^{(-)}(t_1)\hat{E}^{(-)}(t_2)\ket{0}\bra{0}\hat{E}^{(+)}(t_3)\hat{E}^{(+)}(t_4)\ket{\Psi}.
     \label{eq:CF_appendD}
 \end{equation}

Writing the temporal component of the electric field operator in Eq.~\eqref{eq:quantized_EM5B}, the inner products can be computed as follows:
\begin{equation}
    \bra{\Psi}\hat{E}^{(-)}(t_1)\hat{E}^{(-)}(t_2)\ket{0} = -\left( \frac{L_0}{2\pi} \right)^2\int_{-\infty}^{\infty}\mathrm{d}\omega_s'\int_{-\infty}^{\infty}\mathrm{d}\omega_i'\bra{\Psi}\hat{a}^{\dagger}(\omega_s')\hat{a}^{\dagger}(\omega_i')\ket{0}e^{i(\omega_s't_1 + \omega_i't_2)}.
    \label{eq:A5}
\end{equation}

The other inner product is just the complex conjugate of this expression, but evaluated in $t_3$ and $t_4$. Now, using the expression obtained for the biphoton state $\ket{\Psi}$ in Eq.~\eqref{eq:Normalized_bs}, it can be noted that
\begin{align}
    \bra{\Psi}\hat{a}^{\dagger}(\omega_s')\hat{a}^{\dagger}(\omega_i')\ket{0} &= \varepsilon^*\int_{-\infty}^{\infty}\mathrm{d}\omega_s\int_{-\infty}^{\infty}\mathrm{d}\omega_i\bra{0}\hat{a}(\omega_s)\hat{a}(\omega_i)\hat{a}^{\dagger}(\omega_s')\hat{a}^{\dagger}(\omega_i')\ket{0}\Phi^*(\omega_s, \omega_i), \\
    &= \varepsilon^*\bra{0}\hat{J}\ket{0},
    \label{eq:sym_step1}
\end{align}
where
\begin{align}
    \hat{J} &= \int_{-\infty}^{\infty}\mathrm{d}\omega_s\int_{-\infty}^{\infty}\mathrm{d}\omega_i\hat{a}(\omega_i)\hat{a}(\omega_s)\hat{a}^{\dagger}(\omega_s')\hat{a}^{\dagger}(\omega_i')\Phi^*(\omega_s, \omega_i).
\end{align}

To obtain an explicit expression for $\hat{J}\ket{0}$, we use the commutation relation Eq.~\eqref{eq:commute_rel} and follow a similar procedure as in the previous section. With this, we obtain
\begin{equation}
    \hat{J}\ket{0} = (2\pi)^2(\Phi^*(\omega_s',\omega_i') + \Phi^*(\omega_i',\omega_s'))\ket{0}.
\end{equation}

From this equation, we can define the symmetrized JSA function $\tilde{\Phi}(\omega_s',\omega_i')$ as
\begin{equation}
    \tilde{\Phi}(\omega_s', \omega_i') =  \frac{\Phi(\omega_s',\omega_i') + \Phi(\omega_i',\omega_s')}{2},
    \label{eq:sym_JSA}
\end{equation}
such that
\begin{equation}
    \hat{J}\ket{0} = 2(2\pi)^2\tilde{\Phi}^*(\omega_s',\omega_i')\ket{0}.
\end{equation}

Substituting this into Eq.~\eqref{eq:sym_step1}, it is obtained that
\begin{equation}
    \bra{\Psi}\hat{a}^{\dagger}(\omega_s')\hat{a}^{\dagger}(\omega_i')\ket{0} = 2(2\pi)^2\varepsilon^*\tilde{\Phi}^*(\omega_s',\omega_i').
    \label{eq:in_pro*}
\end{equation}

Following the same procedure for the complex conjugate term, but evaluated in $\omega_s$ and $\omega_i$, we obtain the expression,
\begin{equation}
    \bra{0}\hat{a}(\omega_i)\hat{a}(\omega_s)\ket{\Psi} = 2(2\pi)^2\varepsilon\tilde{\Phi}(\omega_s,\omega_i).
    \label{eq:in_pro}
\end{equation}

Replacing Eqs.~\eqref{eq:in_pro*} and~\eqref{eq:in_pro} in Eq.~\eqref{eq:A5} and taking into account that the complex conjugate is evaluated in $t_3$ and $t_4$, the field correlation function is 
\begin{equation}
    C_F(t_1,t_2,t_3,t_4) = 4L_0^4\int_{-\infty}^{\infty}\mathrm{d}\omega_s'\int_{-\infty}^{\infty}\mathrm{d}\omega_i'\int_{-\infty}^{\infty}\mathrm{d}\omega_s\int_{-\infty}^{\infty}\mathrm{d}\omega_i|\varepsilon|^2\tilde{\Phi}^{*}(\omega_s',\omega_i')\tilde{\Phi}(\omega_s,\omega_i)e^{i(\omega_s't_1+\omega_i't_2-\omega_st_3-\omega_it_4)}.
    \label{eq:CF_A}
\end{equation}

Now, it must hold that $\omega_s' + \omega_i' = \omega_s + \omega_i$, therefore, $\omega_i' = \omega_s + \omega_i - \omega_s'$, and
\begin{equation}
    \small
    e^{i(\omega_s't_1 + \omega_i't_2)}e^{-i(\omega_st_3 + \omega_it_4)} = e^{i(\omega_s't_1 - \omega_s't_2 + \omega_st_2 + \omega_it_2)}e^{-i(\omega_st_3 + \omega_it_4)}.
\end{equation}

Introducing the change of variables $\tau = t_2-t_1$, $s=t_3-t_2$ and $r=t_4-t_3$ and rearranging terms in the previous equation, it follows that
\begin{equation}
    e^{i(\omega_s't_1 + \omega_i't_2)}e^{-i(\omega_st_3 + \omega_it_4)} = e^{-i\omega_s'\tau}e^{-i(\omega_s + \omega_i)s}e^{-i\omega_ir}.
\end{equation}

Therefore
\begin{equation}
    C_F = \kappa\int_{-\infty}^{\infty}\mathrm{d}\omega_s'\int_{-\infty}^{\infty}\mathrm{d}\omega_i'\int_{-\infty}^{\infty}\mathrm{d}\omega_s\int_{-\infty}^{\infty}\mathrm{d}\omega_i|\varepsilon|^2\tilde{\Phi}^{*}(\omega_s',\omega_i')\tilde{\Phi}(\omega_s,\omega_i)e^{-i\omega_s'\tau}e^{-(\omega_s + \omega_i)s}e^{-i\omega_ir},
\end{equation}
where $\kappa=\left(\frac{\hslash\omega_0}{\epsilon_0cA_0}\right)^2=4L_0^4$. Also, we can define the flux of photon pairs in a time $T$ produced by the SPDC process as $\phi=2|\varepsilon|^2/A_0T$, where the factor of $2$ comes from the fact that we have photon pairs.

\section{ETPA Cross-Section}
\label{append:E}
\subsection{ETPA probability}

The ETPA probability in Eq.~\eqref{eq:TPAprobability1} can be written as
\begin{equation}
    p_{g\rightarrow f}=\frac{1}{\hslash^{4}}\int_{-\infty}^{t}dt_4\int_{-\infty}^{t_4}dt_3\int_{-\infty}^{t_3}dt_2\int_{-\infty}^{t_2}dt_1 C_F(t_1,t_2,t_3,t_4)C_A(t_1,t_2,t_3,t_4) + c.c.. 
    \label{eq:ETPA_prop_A}
\end{equation}

The time integrals are performed considering the expressions for the correlation functions in Eqs.~\eqref{eq:atomcorrelation} and \eqref{eq:fieldcorrelation2}. The ETPA probability can be expressed as follows considering the change of variables $\tau = t_2-t_1$, $s=t_3-t_2$, $r=t_4-t_3$ and $\omega_i' = \omega_s + \omega_i - \omega_s'$:
\begin{align*}
    p_{g\to f} &= \frac{|\varepsilon|^2\kappa}{\hslash^4}\sum_{e,e'}D_{fe'eg}\int_{-\infty}^{\infty}\mathrm{d}\omega_s'\int_{-\infty}^{\infty}\mathrm{d}\omega_s\int_{-\infty}^{\infty}\mathrm{d}\omega_i \\
    &\times \int_{0}^{\infty}\mathrm{d}re^{-(\gamma_{fe'}-i\omega_{fe'}+i\omega_i)r}\int_{0}^{\infty}\mathrm{d}se^{-(\gamma_{fg}-i\omega_{fg}+i\omega_s + i\omega_i)s}\int_{0}^{\infty}\mathrm{d}\tau e^{-(\gamma_{eg}-i\omega_{eg}+i\omega_s')\tau}\Phi^*(\omega_s',\omega_i')\Phi(\omega_s,\omega_i),
\end{align*}

Defining
\begin{align*}
    I_{e,e'} &=\int_{-\infty}^{\infty}\mathrm{d}\omega_s'\int_{-\infty}^{\infty}\mathrm{d}\omega_s\int_{-\infty}^{\infty}\mathrm{d}\omega_i \\
    &\times \int_{0}^{\infty}\mathrm{d}re^{-(\gamma_{fe'}-i\omega_{fe'}+i\omega_i)r}\int_{0}^{\infty}\mathrm{d}se^{-(\gamma_{fg}-i\omega_{fg}+i\omega_s + i\omega_i)s}\int_{0}^{\infty}\mathrm{d}\tau e^{-(\gamma_{eg}-i\omega_{eg}+i\omega_s')\tau}\tilde{\Phi}^*(\omega_s',\omega_i')\tilde{\Phi}(\omega_s,\omega_i),
\end{align*}
and performing the integrations with respect to $r$, $s$ and $\tau$, it is obtained that
\begin{equation}
    I_{e,e'} = \int_{-\infty}^{\infty}\mathrm{d}\omega_s'\int_{-\infty}^{\infty}\mathrm{d}\omega_s\int_{-\infty}^{\infty}\mathrm{d}\omega_i\frac{\tilde{\Phi}^{*}(\omega_s',\omega_i')\tilde{\Phi}(\omega_s,\omega_i)}{[\gamma_{fe'}+i(\omega_i - \omega_{fe'})]\left[\gamma_{fg}+i(\omega_s+\omega_i-\omega_{fg})\right][\gamma_{eg} +i(\omega_s' - \omega_{eg})]},
    \label{eq:brrr4}
\end{equation}
and the ETPA probability is therefore
\begin{equation}
	p_{g\rightarrow f} =\frac{|\varepsilon|^2\kappa}{\hslash^{4}}\sum_{e,e'}D_{fe'eg}I_{e,e'}+ c.c.
	\label{eq:TPAprobability_A}
\end{equation}

\subsection{ETPA Cross-Section in the Far-Off-Resonance Approximation}
\label{append:H}

In this section we compute the integrals in Eq.~\eqref{eq:brrr4} considering the Far-Off-Resonance approximation, therefore,
\begin{equation}
    I_{e,e'} = \frac{1}{(-\omega_{fe'}+\omega_0)(\omega_{eg}-\omega_0)}\int_{-\infty}^{\infty}d\omega_s'\int_{-\infty}^{\infty}d\omega_s\int_{-\infty}^{\infty}d\omega_i\frac{\tilde{\Phi}^{*}(\omega_s',\omega_i')\tilde{\Phi}(\omega_s,\omega_i)}{\gamma_{fg}-i\omega_{fg}+i\omega_s+i\omega_i} + c.c.
    \label{eq:integralesH}
\end{equation}

Considering the sum with complex conjugate term and the fact that $\omega_p=\omega_s+\omega_i=\omega_s'+\omega_i'$, this expression can be rewritten as
\begin{equation}
    I_{e,e'} = \frac{2\gamma_{fg}}{(-\omega_{fe'}+\omega_0)(\omega_{eg}-\omega_0)}\int_{-\infty}^{\infty}d\omega_s'\int_{-\infty}^{\infty}d\omega_s\int_{-\infty}^{\infty}d\omega_p\frac{\tilde{\Phi}^{*}(\omega_s',\omega_p - \omega_s')\tilde{\Phi}(\omega_s,\omega_p - \omega_s)}{\gamma_{fg}^2 + (\omega_p-\omega_{fg})^2}.
    \label{eq:integrales2H}
\end{equation}

In order to simplify the integral, it is convenient to write it in terms of the detunings defined in Eq.~\eqref{eq:detunings}, so that
\begin{equation}
    I_{e,e'} = \frac{2\gamma_{fg}}{(-\omega_{fe'}+\omega_0)(\omega_{eg}-\omega_0)}\int_{-\infty}^{\infty}d\Omega_s'\int_{-\infty}^{\infty}d\Omega_s\int_{-\infty}^{\infty}d\Omega_p\frac{\tilde{\Phi}^{*}(\Omega_s'+\omega_0,\omega_0+\Omega_p - \Omega_s')\tilde{\Phi}(\Omega_s+\omega_0,\omega_0+\Omega_p - \Omega_s)}{\gamma_{fg}^2 + (\Omega_p + 2\omega_0-\omega_{fg})^2}.
\end{equation}

The entangled-photon pairs interacting with the atom have the JSA in Eq.~\eqref{eq:type-II-JSA1}, along with Eq.~\eqref{eq:delta}: 
\begin{equation}
    I_{e,e'} = K_{e,e'}\gamma_{fg}\int_{-\infty}^{\infty}d\Omega_s'\int_{-\infty}^{\infty}d\Omega_s\int_{-\infty}^{\infty}d\Omega_p
    \delta(\Omega_p)\frac{\tilde{\varphi}^*(\Omega_s',\Omega_p - \Omega_s') 
    \tilde{\varphi}(\Omega_s,\Omega_p - \Omega_s)}{\gamma_{fg}^2+(\Omega_p + 2\omega_0-\omega_{fg})^2},
\end{equation}
where
\begin{equation}
    K_{e,e'} = \frac{T_e/\pi}{(-\omega_{fe'}+\omega_0)(\omega_{eg}-\omega_0)}.
\end{equation}

Integrating over $\Omega_p$ results in $\Omega_p=0$ and $\Omega_s=\Omega_i$. Therefore this expression becomes
\begin{equation}
    I_{e,e'} = K_{e,e'}\gamma_{fg}\int_{-\infty}^{\infty}d\Omega_s'\int_{-\infty}^{\infty}d\Omega_s\frac{\text{sinc}(\Omega_s' T_e/2)
    \text{sinc}(\Omega_s T_e/2)}{\gamma_{fg}^2+(2\omega_0 -\omega_{fg})^2},
\end{equation}

Using the relation $\Delta = 2\omega_0 - \omega_{fg}$, this equation can be simplified to
\begin{equation}
    I_{e,e'} = K_{e,e'}\frac{\gamma_{fg}}{\gamma_{fg}^2 + \Delta^2}\int_{-\infty}^{\infty}d\Omega_s'\text{sinc}\left(\frac{\Omega_s' T_e}{2} \right)\int_{-\infty}^{\infty}d\Omega_s\text{sinc}\left(\frac{\Omega_s T_e}{2} \right).
\end{equation}

The integral of the cardinal sine function is a very well-known result:
\begin{equation}
    \int_{-\infty}^{\infty}d\Omega_s\text{sinc}\left(\frac{\Omega_s T_e}{2} \right) = \frac{2\pi}{T_e},
\end{equation}
therefore, the $I_{e,e'}$ can be written as
\begin{equation}
    I_{e,e'} = \left( \frac{2\pi}{T_e} \right)^2\frac{\gamma_{fg}}{\gamma_{fg}^2 + \Delta^2} K_{e,e'},
\end{equation}
or
\begin{equation}
    I_{e,e'} = \frac{4\pi}{T_e}\frac{1}{(-\omega_{fe'}+\omega_0)(\omega_{eg}-\omega_0)}\frac{\gamma_{fg}}{\gamma_{fg}^2 + \Delta^2},
\end{equation}
and the ETPA cross section in Eq.~\eqref{eq:sigmae_prime} takes the form
\begin{equation}
    \sigma_e^{(2)} = \frac{2\pi A_0\kappa}{T_e\hslash^4}\ \sum_{e,e'}\frac{D_{fe'eg}}{(-\omega_{fe'}+\omega_0)(\omega_{eg}-\omega_0)}\frac{\gamma_{fg}}{\gamma_{fg}^2 + \Delta^2}.
\end{equation}

\subsection{ETPA Cross Section without the Far-Off resonance approximation}
\label{append:G}
In this section we compute integrals in Eq.~\eqref{eq:brrr4} such that $I_{e,e'}$ is
\begin{equation}
    \small
    I_{e,e'} = \int_{-\infty}^{\infty}\mathrm{d}\omega_s'\int_{-\infty}^{\infty}\mathrm{d}\omega_s\int_{-\infty}^{\infty}\mathrm{d}\omega_p\frac{\tilde{\Phi}^{*}(\omega_s',\omega_p - \omega_s')\tilde{\Phi}(\omega_s,\omega_p - \omega_s)}{[\gamma_{fe'}+i(\omega_p - \omega_s - \omega_{fe'})][\gamma_{fg}+i(\omega_p - \omega_{fg})][\gamma_{eg} +i(\omega_s' - \omega_{eg})]} + c.c.
    \label{eq:brrr6}
\end{equation}

In order to simplify the integral, we write it in terms of the detunings defined in  Eq.~\eqref{eq:detunings}. We also introduce the detunings of the central frequency of the down-converted photons with respect to the transition frequencies to the intermediate states, $\Omega_{fe'} = \omega_0 - \omega_{fe'}$ and $\Omega_{eg} = \omega_0 - \omega_{eg}$ and the detuning of the pump central frequency with respect to the two-photon transition frequency, $\Delta = 2\omega_0 - \omega_{fg}$. With this,  $I_{e,e'}$ can be expressed as
\begin{align}
    I_{e,e'} &= \int_{-\infty}^{\infty}\mathrm{d}\Omega_s'\int_{-\infty}^{\infty}\mathrm{d}\Omega_s\int_{-\infty}^{\infty}\mathrm{d}\Omega_p \notag \\
    &\times\frac{\tilde{\Phi}^{*}(\Omega_s'+\omega_0,\omega_0 +\Omega_p - \Omega_s')\tilde{\Phi}(\Omega_s+\omega_0,\omega_0 +\Omega_p - \Omega_s)}{[\gamma_{fe'}+i(\Omega_p  - \Omega_s+ \Omega_{fe'})][\gamma_{fg}+i(\Omega_p + \Delta)][\gamma_{eg} +i(\Omega_s' + \Omega_{eg})]} + c.c.
    \label{eq:brrr7}
\end{align}

The entangled-photon pairs interacting with the atom have the JSA in Eq.~\eqref{eq:type-II-JSA1}, along with Eq.~\eqref{eq:delta}  and
\begin{align}
    I_{e,e'} &= \int_{-\infty}^{\infty}\mathrm{d}\Omega_s'\int_{-\infty}^{\infty}\mathrm{d}\Omega_s\int_{-\infty}^{\infty}\mathrm{d}\Omega_p\delta(\Omega_p) \notag \\
    &\times\frac{\tilde{\varphi}^{*}(\Omega_s',\Omega_p - \Omega_s')\tilde{\varphi}(\Omega_s,\Omega_p - \Omega_s)}{[\gamma_{fe'}+i(\Omega_p  - \Omega_s + \Omega_{fe'})][\gamma_{fg}+i(\Omega_p + \Delta)][\gamma_{eg} +i(\Omega_s' + \Omega_{eg})]} + c.c.
    \label{eq:brrr8}
\end{align}

Integrating over $\Omega_p$ results in $\Omega_p=0$ and $\Omega_s=\Omega_i$. Therefore, 
\begin{equation}
    \small
    I_{e,e'} = \int_{-\infty}^{\infty}\mathrm{d}\Omega_s'\int_{-\infty}^{\infty}\mathrm{d}\Omega_s\frac{\tilde{\varphi}^*(\Omega_s')\tilde{\varphi}(\Omega_s)}{[\gamma_{fe'}+i(\Omega_{fe'} - \Omega_s)][\gamma_{fg}+i(2\omega_0 - \omega_{fg})][\gamma_{eg} +i(\Omega_s' + \Omega_{eg})]} + c.c.
\end{equation}

Writing the symmetrized JSA function explicitly from Eq.~\eqref{eq:type-II-JSA1} and Eq.~\eqref{eq:sym_JSA}, we get that
\begin{equation}
    \small
    I_{e,e'} = \frac{T_e}{2\pi}\int_{-\infty}^{\infty}\mathrm{d}\Omega_s'\int_{-\infty}^{\infty}\mathrm{d}\Omega_s\frac{\text{sinc}(\Omega_s' T_e/2)
    \text{sinc}(\Omega_s T_e/2)}{[\gamma_{fe'}+i(\Omega_{fe'} - \Omega_s)](\gamma_{fg}+i\Delta)[\gamma_{eg} +i(\Omega_s' + \Omega_{eg})]} + c.c.,
\end{equation}

We simplify the computation of the sum with the complex conjugate term by introducing the change of variables $\Omega = \Omega_s - \Omega_{fe'}$, $\Omega' = \Omega_s' + \Omega_{eg}$, such that
\begin{equation}
    \frac{1}{(\gamma_{fe'} - i\Omega)(\gamma_{fg} + i\Delta)(\gamma_{eg} + i\Omega')} + c.c = \frac{2(\gamma_{fe'}\gamma_{fg}\gamma_{eg} + \gamma_{eg}\Omega\Delta - \gamma_{fe'}\Omega' \Delta+ \gamma_{fg}\Omega\Omega')}{(\gamma_{fe'}^2 + \Omega^2)(\gamma_{fg}^2 + \Delta^2)(\gamma_{eg}^2 + \Omega'^2)},
\end{equation}
and
\begin{equation}
    \small
    I_{e,e'} = 2\frac{T_e}{2\pi}\int_{-\infty}^{\infty}\mathrm{d}\Omega'\int_{-\infty}^{\infty}\mathrm{d}\Omega\text{sinc}[(\Omega' - \Omega_{eg})T_e/2]
    \text{sinc}[(\Omega + \Omega_{fe'})T_e/2]\frac{(\gamma_{fe'}\gamma_{fg}\gamma_{eg} + \gamma_{eg}\Omega\Delta - \gamma_{fe'}\Omega'\Delta + \gamma_{fg}\Omega\Omega')}{(\gamma_{fe'}^2 + \Omega^2)(\gamma_{fg}^2 + \Delta^2)(\gamma_{eg}^2 + \Omega'^2)},
    \label{eq:brrr9}
\end{equation}

Notice that the four terms in the integral can be expressed in terms of the following two functions
\begin{equation}
    S_1(\Omega_{jk}, \gamma_{jk}, T_e) = \int_{-\infty}^{\infty}\frac{\text{sinc}[(\Omega + \Omega_{jk})T_e/2]}{(\gamma_{jk}^2 + \Omega^2)}\mathrm{d}\Omega,
\end{equation}
and
\begin{equation}
    S_2(\Omega_{jk}, \gamma_{jk}, T_e) = \int_{-\infty}^{\infty}\frac{\Omega\text{sinc}[(\Omega + \Omega_{jk})T_e/2]}{(\gamma_{jk}^2 + \Omega^2)}\mathrm{d}\Omega,
\end{equation}
with $\Omega_{jk}=\Omega_{fe'}$ or $\Omega_{jk}=-\Omega_{eg}$. Explicitly  we separate Eq.~\eqref{eq:brrr9} in the four terms
\begin{equation}
    I_{e,e'} = 2\frac{T_e}{2\pi}(I_{e,e'}^{(1)} + I_{e,e'}^{(2)} + I_{e,e'}^{(3)} + I_{e,e'}^{(4)}),
    \label{eq:I}
\end{equation}
where

\begin{align}
    \label{eq:R1}
    I_{e,e'}^{(1)} &= \frac{\gamma_{fe'}\gamma_{fg}\gamma_{eg}}{\gamma_{fg}^2 + \Delta^2}S_1(\Omega_{fe'}, \gamma_{fe'}, T_e) S_1(-\Omega_{eg}, \gamma_{eg}, T_e), \\
    \label{eq:R2}
    I_{e,e'}^{(2)} &= \frac{\gamma_{eg}\Delta}{\gamma_{fg}^2 + \Delta^2} S_2(\Omega_{fe'}, \gamma_{fe'}, T_e) S_1(-\Omega_{eg}, \gamma_{eg}, T_e), \\
    \label{eq:R3}
    I_{e,e'}^{(3)} &= -\frac{\gamma_{fe'}\Delta}{\gamma_{fg}^2 + \Delta^2}S_1(\Omega_{fe'}, \gamma_{fe'}, T_e)S_2(-\Omega_{eg}, \gamma_{eg}, T_e), \\
    \label{eq:R4}
    I_{e,e'}^{(4)} &= \frac{\gamma_{fg}}{\gamma_{fg}^2 + \Delta^2} S_2(\Omega_{fe'}, \gamma_{fe'}, T_e)S_2(-\Omega_{eg}, \gamma_{eg}, T_e).
\end{align}

We compute each term individually, starting with $I_{e,e'}^{(1)}$. To calculate this term, we need to perform the integrals of the $S_1$ and $S_2$ functions. The former can be calculated as:
\begin{equation}
    S_1(\Omega_{fe'}, \gamma_{fe'}, T_e)= \frac{2}{T_e}\text{Im}\left[ \int_{-\infty}^{\infty}\frac{\text{exp}\left[ i(\Omega + \Omega_{fe'})T_e/2 \right]}{(\Omega + \Omega_{fe'})(\gamma_{fe'}^2 + \Omega^2)}\mathrm{d}\Omega \right] = \frac{2}{T_e}\text{Im}\left[ \int_{-\infty}^{\infty}F(\Omega)\mathrm{d}\Omega \right].
    \label{eq:int_1}
\end{equation}

\begin{figure}[h!]
	\centering
	\includegraphics[width=0.7\linewidth]{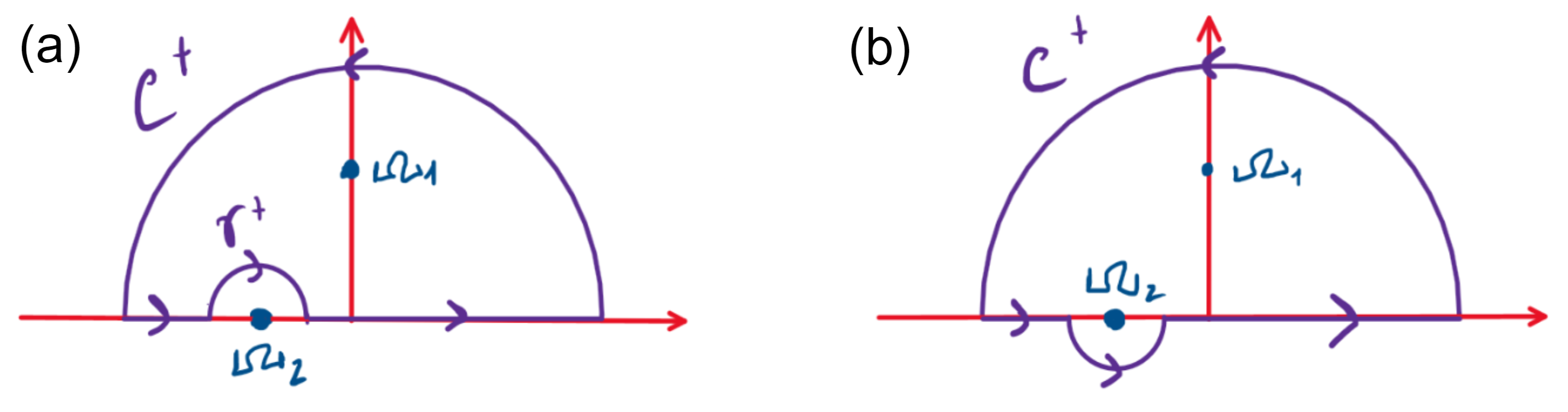}
	\caption[Schematic representation of both poles of the function $S(\Omega)$ in the complex plane for $\Omega$]{Schematic representation of both poles of the function $F(\Omega)$ in the complex plane for $\Omega$. The contour of a) removes the real pole $\Omega_2$ from the enclosed region. The contour of b) encloses the real pole $\Omega_2$ in the region of interest.}
	\label{fig:complex_plane_2}
\end{figure}

As shown in Fig.~\ref{fig:complex_plane_2}, the function $F(\Omega)$ has poles at $\Omega_1 = i\gamma_{fe'}$ (upper half-plane) and $\Omega_2 = -\Omega_{fe'}$ (real axis). Using the residue theorem with appropriate treatment of the real pole through the principal value integration, we obtain:
\begin{equation}
	\int_{-\infty}^{\infty}F(\Omega)\mathrm{d}\Omega = 2\pi i \left( a_{-1}^{(\Omega_1)} + \frac{a_{-1}^{(\Omega_2)}}{2} \right),
	\label{eq:S_integral}
\end{equation}

where $a_{-1}^{(\Omega_{1,2})}$ denotes the residue at the pole $\Omega_{1,2}$. The first residue $a_{-1}^{(\Omega_1)}$ is evaluated to be
\begin{equation}
    a_{-1}^{(\Omega_1)} = -\frac{e^{-(\gamma_{fe'} - i\Omega_{fe'})T_e/2}}{2\gamma_{fe'}(\gamma_{fe'} - i\Omega_{fe'})},
    \label{eq:residue_1}
\end{equation}
while the second residue $a_{-1}^{(\Omega_2)}$ is found to be
\begin{equation}
    a_{-1}^{(\Omega_2)} = \frac{1}{\gamma_{fe'}^2 + \Omega_{fe'}^2}.
    \label{eq:residue_2}
\end{equation}

Substituting Eqs.~\eqref{eq:residue_1} and \eqref{eq:residue_2} in Eq.~\eqref{eq:S_integral}, and taking the imaginary part of the resulting expression, it follows that
\begin{equation}
    S_1(\Omega_{fe'}, \gamma_{fe'}, T_e) = \frac{2\pi \gamma_{fe'} -2\pi  e^{-\gamma_{fe'}T_e/2} \left[ \gamma_{fe'} \cos \left(\frac{\Omega_{fe'}T_e}{2}\right)-\Omega_{fe'} \sin \left(\frac{ \Omega_{fe'}T_e}{2}\right)\right]}{T_e\gamma_{fe'} \left(\gamma_{fe'}^2+\Omega_{fe'}^2\right)}.
    \label{eq:int_3}
\end{equation}

We continue with the second term in Eq.~\eqref{eq:I}, $I_{e,e'}^{(2)}$. We  need to evaluate the integral $S_2(\Omega_{fe'}, \gamma_{fe'}, T_e)$.  This integral can be computed, similarly to Eq.~\eqref{eq:int_1}, as 
\begin{equation}
     S_2(\Omega_{fe'}, \gamma_{fe'}, T_e) = \frac{2}{T_e}\text{Im}\left[ \int_{-\infty}^{\infty}\frac{\Omega \text{exp}\left[ i(\Omega+\Omega_{fe'})T_e/2 \right]}{(\Omega+\Omega_{fe'})(\gamma_{fe'}^2 + \Omega^2)}\mathrm{d}\Omega \right].
\end{equation} 

Following the same procedure as for the integral in Eq.~\eqref{eq:int_1}, the result will correspond to the same as in Eq.~\eqref{eq:S_integral}:
\begin{equation}
	\int_{-\infty}^{\infty}\frac{\Omega \text{exp}\left[ i(\Omega+\Omega_{fe'})T_e/2 \right]}{(\Omega+\Omega_{fe'})(\gamma_{fe'}^2 + \Omega^2)}\mathrm{d}\Omega = 2\pi i \left( a_{-1}^{(\Omega_1)} + \frac{a_{-1}^{(\Omega_2)}}{2} \right),
	\label{eq:S_integral2}
\end{equation}
with the poles $\Omega_1$ and $\Omega_2$ being the same, but with the residues being computed as
\begin{equation}
    a_{-1}^{(\Omega_1)} = \frac{e^{-(\gamma_{fe'} - i\Omega_{fe'})T_e/2}}{2(i\gamma_{fe'} + \Omega_{fe'})},
    \label{eq:residue_3}
\end{equation}
and
\begin{equation}
    a_{-1}^{(\Omega_2)} = -\frac{\Omega_{fe'}}{\gamma_{fe'}^2 + \Omega_{fe'}^2}.
    \label{eq:residue_4}
\end{equation}

Substituting Eqs.~\eqref{eq:residue_3} and \eqref{eq:residue_4} into Eq.~\eqref{eq:S_integral2} and taking the imaginary part of the resulting expression, we obtain that
\begin{align}
    S_2(\Omega_{fe'}, \gamma_{fe'}, T_e) = \frac{-2\pi\Omega_{fe'} +2\pi  e^{-\gamma_{fe'}T_e/2} \left[ \gamma_{fe'} \sin \left(\frac{\Omega_{fe'}T_e}{2}\right)+\Omega_{fe'} \cos \left(\frac{ \Omega_{fe'}T_e}{2}\right)\right]}{T_e \left(\gamma_{fe'}^2+\Omega_{fe'}^2\right)}.
    \label{eq:int_6}
\end{align}

With this we can write the explicit form of the the functions $I_{e,e'}$

\begin{align}
    \label{eq:R1}
    I_{e,e'}^{(1)} &= \frac{\gamma_{fe'}\gamma_{fg}\gamma_{eg}}{\gamma_{fg}^2 + \Delta^2}S_1(\Omega_{fe'}, \gamma_{fe'}, T_e) S_1(-\Omega_{eg}, \gamma_{eg}, T_e), \\
    \label{eq:R2}
    I_{e,e'}^{(2)} &= \frac{\gamma_{eg}\Delta}{\gamma_{fg}^2 + \Delta^2} S_2(\Omega_{fe'}, \gamma_{fe'}, T_e) S_1(-\Omega_{eg}, \gamma_{eg}, T_e), \\
    \label{eq:R3}
    I_{e,e'}^{(3)} &= -\frac{\gamma_{fe'}\Delta}{\gamma_{fg}^2 + \Delta^2}S_1(\Omega_{fe'}, \gamma_{fe'}, T_e)S_2(-\Omega_{eg}, \gamma_{eg}, T_e), \\
    \label{eq:R4}
    I_{e,e'}^{(4)} &= \frac{\gamma_{fg}}{\gamma_{fg}^2 + \Delta^2} S_2(\Omega_{fe'}, \gamma_{fe'}, T_e)S_2(-\Omega_{eg}, \gamma_{eg}, T_e).
\end{align}
and the ETPA cross section in Eq.~\eqref{eq:sigmae_prime} is
\begin{equation}
    \sigma_e^{(2)} = \frac{A_0\kappa}{\hslash^4}\frac{T_e}{2\pi}\sum_{e,e'}D_{fe'eg}(I_{e,e'}^{(1)}+I_{e,e'}^{(2)}+I_{e,e'}^{(3)}+I_{e,e'}^{(4)}).
\end{equation}

\section{Atomic structure information for cesium}
\label{sec:cs_values}

\begin{figure}[ht]
    \centering
    \includegraphics[width = 0.6 \linewidth]{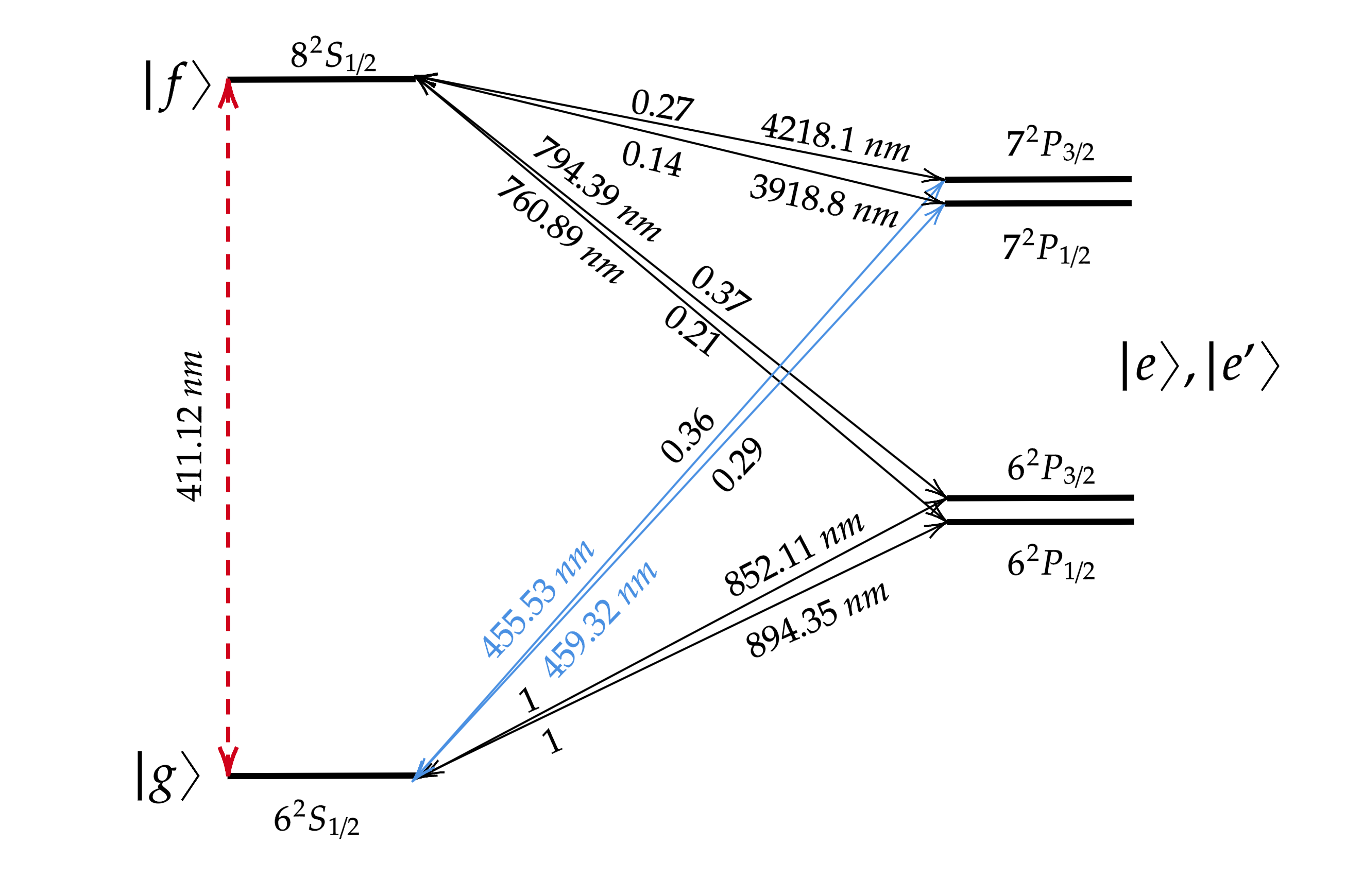}
    \caption[Energy levels of atomic cesium]{Energy levels of the cesium atom relevant for this work. Solid lines represent one-photon transitions. The dashed line corresponds to the two-photon transition. The wavelengths and branching rations of each transition are shown in the figure. Image modified from \cite{CaracasNunez:23}.}
    \label{fig:cstructure}
\end{figure}

The energy levels and the spectroscopic information for Cs are presented in Fig. \ref{fig:cstructure} and Table \ref{tab:Dipole}. This information is employed in the calculation of the ETPA cross section.  

\begin{table}[ht]
\centering
\small
\begin{tabular}{|c c c c c c c|}
\hline
\textbf{Transition} & \textbf{\begin{tabular}[c]{@{}c@{}}Frequency \\ $[$THz$]$ \end{tabular}} &  $d_{ij}$ [$10^{-10}$ m] & \textbf{\begin{tabular}[c]{@{}c@{}}Branching\\ Ratio\end{tabular}} & \textbf{Ref.}  &\ \textbf{\begin{tabular}[c]{@{}c@{}}Decay Rate \\ $\times 10^{6} [rad/s]$ \end{tabular}} & \textbf{Ref.} \\ \hline
$6S_{1/2}\rightarrow 6P_{1/2}$ & 335.116048807(41) & 1.6840   & 1  & \cite{Udem1998,steck_1998} & $2\pi\cdot 4.5612$ & \cite{steck_1998} \\ \hline
$6S_{1/2}\rightarrow 6P_{3/2}$ & 351.72571850(11) &   2.3700 &  1 & \cite{Udem2000,steck_1998} & $2\pi\cdot 5.2227$ & \cite{steck_1998} \\ \hline
$6S_{1/2}\rightarrow 7P_{1/2}$ & 652.50461 &  0.14859 &  0.29  & \cite{Sansonetti2009,Antypas2013} & $6.0529$ & \cite{toh2019measurement} \\ \hline
$6S_{1/2}\rightarrow 7P_{3/2}$ & 657.93241   & 0.30586 &  0.36 & \cite{Sansonetti2009,Antypas2013} & $7.2706$ & \cite{toh2019measurement} \\ \hline
$6P_{1/2}\rightarrow 8S_{1/2}$ & 393.89197 & 0.5429  & 0.21 & \cite{Sansonetti2009,Gunawardena2007} & $2.0202$ & \cite{horbatsch2005classical} \\ \hline
$6P_{3/2}\rightarrow 8S_{1/2}$ & 377.28410 &  0.7731  & 0.37 & \cite{Sansonetti2009,Gunawardena2007} & $3.5594$ & \cite{horbatsch2005classical} \\ \hline
$7P_{1/2}\rightarrow 8S_{1/2}$ & 76.501147 & 4.928  & 0.14 & \cite{Sansonetti2009, Gunawardena2007} & $1.3468$ & \cite{horbatsch2005classical} \\ \hline
$7P_{3/2}\rightarrow 8S_{1/2}$ & 71.0774$^{a}$ &  7.443  & 0.27& \cite{Gunawardena2007} & $2.5974$ & \cite{horbatsch2005classical} \\ \hline
$6S_{1/2}\rightarrow 8S_{1/2}$ & 729.009799  & TPA & TPA & \cite{Sansonetti2009} & $9.6200$ & \cite{ALESSANDRETTI1977289} \\ 
$F_i=3\rightarrow F_f=3$ & 729.014477  & TPA  & TPA & \cite{Wu} & &\\ 
$F_i=4\rightarrow F_f=4$ & 729.006161 & TPA  & TPA & \cite{Wu} & & \\ \hline
\end{tabular}
\caption{Summary of the transition frequencies, $\nu$, dipole moments per unit charge, $d_{ij}$, and branching ratios relevant for the theoretical study of $\sigma$ with the respective references. $^{a}$ The frequency value of this transition is obtained from the difference between the frequencies $\nu_{6S_{1/2}\rightarrow 8S_{1/2}}$ and $\nu_{6S_{1/2}\rightarrow 7P_{3/2}}$. Table adapted from \cite{CaracasNunez:23}.}
\label{tab:Dipole}
\end{table}

\nocite{*}

\bibliography{bibliography}

\end{document}